\documentclass[journal=jacsat,manuscript=article]{achemso}

\SectionNumbersOn
\usepackage[version=3]{mhchem} 
\usepackage[dvipsnames,table]{xcolor} 
\usepackage{amsmath} 
\usepackage{chemformula} 
\usepackage[export]{adjustbox}
\usepackage{float}
\usepackage{lastpage}
\usepackage{fancyhdr}
\usepackage{caption}

\usepackage{xr}
\makeatletter

\usepackage[T1]{fontenc}

\newcommand*{\addFileDependency}[1]{
\typeout{(#1)}
\@addtofilelist{#1}
\IfFileExists{#1}{}{\typeout{No file #1.}}
}\makeatother

\newcommand*{\myexternaldocument}[1]{%
\externaldocument{#1}%
\addFileDependency{#1.tex}%
\addFileDependency{#1.aux}%
}

\myexternaldocument{supporting-information}

\usepackage[version=3]{mhchem} 
\usepackage[dvipsnames]{xcolor} 
\usepackage{chemformula} 
\usepackage[export]{adjustbox}
\usepackage{float}
\usepackage{graphicx}
\usepackage{booktabs}

\usepackage[T1]{fontenc}
\SectionNumbersOn

\usepackage[T1]{fontenc}

\usepackage{graphicx}
\usepackage{booktabs}

\author{Julia H. Yang}
\affiliation[HUCE]
{Center for the Environment, Harvard University, Cambridge MA 02138}
\alsoaffiliation[SEAS]
{Harvard John A. Paulson School of Engineering and Applied Sciences, Harvard University, Cambridge MA 02138}
\email{jhyang@g.harvard.edu}
\author{Amanda Whai Shin Ooi}
\affiliation[ChemEColumbia]
{Department of Chemical Engineering, Columbia University in the City of New York, New York City, NY 10027}
\author{Zachary A. H. Goodwin}
\affiliation[SEAS]
{Harvard John A. Paulson School of Engineering and Applied Sciences, Harvard University, Cambridge MA 02138}
\author{Yu Xie}
\affiliation[SEAS]
{Harvard John A. Paulson School of Engineering and Applied Sciences, Harvard University, Cambridge MA 02138}
\alsoaffiliation[MSFT]
{Microsoft Research AI for Science, Berlin, Germany 10179}
\author{Jingxuan Ding}
\affiliation[SEAS]
{Harvard John A. Paulson School of Engineering and Applied Sciences, Harvard University, Cambridge MA 02138}
\author{Stefano Falletta}
\affiliation[SEAS]
{Harvard John A. Paulson School of Engineering and Applied Sciences, Harvard University, Cambridge MA 02138}
\author{Ah-Hyung Alissa Park}
\affiliation[UCLA]{Department of Chemical and Biomolecular Engineering, University of California, Los Angeles, Los Angeles, CA 90095}
\author{Boris Kozinsky}
\affiliation[HUCE]
{Center for the Environment, Harvard University, Cambridge MA 02138}
\alsoaffiliation[SEAS]
{Harvard John A. Paulson School of Engineering and Applied Sciences, Harvard University, Cambridge MA 02138}
\alsoaffiliation[Bosch]
{Robert Bosch Research and Technology Center, Watertown MA 02472}

\title[An \textsf{achemso} demo]
  {Room-temperature decomposition of the ethaline deep eutectic solvent}


\begin{document}

\clearpage
\begin{abstract}
Environmentally-benign, non-toxic electrolytes with combinatorial design spaces are excellent candidates for green solvents, green leaching agents, and carbon capture sources. We examine ethaline, a 2:1 molar ratio of ethylene glycol and choline chloride. Despite its touted green credentials, we find partial decomposition of ethaline into toxic chloromethane and dimethylaminoethanol at room temperature, limiting its sustainable advantage. We experimentally characterize these decomposition products and computationally develop a general, quantum chemically-accurate workflow to understand its decomposition. We find that fluctuations of the hydrogen bonds bind chloride near reaction sites, initiating the reaction between choline cations and chloride anions. The strong hydrogen bonds formed in ethaline are resistant to thermal perturbations, entrapping Cl in high-energy states, promoting the uphill reaction. The reaction appears to be unavoidable in ethaline. In the design of stable green solvents, we recommend detailed evaluation of the hydrogen-bonding potential energy landscape as a key consideration for generating stable solvent-mixtures.
   
\end{abstract}

\begin{figure}
    \centering
    \includegraphics[width=4.45cm]{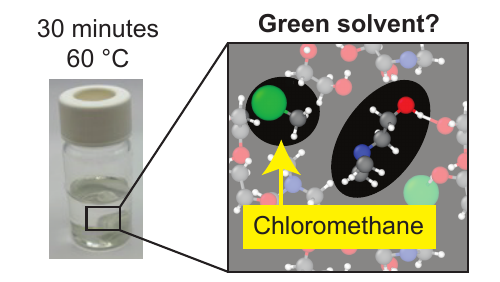}
    \caption*{Table of Contents image}
    \label{fig:toc}
\end{figure}
\clearpage 

The International Energy Agency (IEA) predicts that demand for critical minerals for enabling the clean energy transition will triple by 2030 and quadruple by 2040 in the Net Zero roadmap. These minerals are primarily copper and aluminum for electricity networks; rare earth elements for permanent magnets in wind turbines; and lithium, nickel, and cobalt for electric vehicles (EVs). For EVs, it estimated that by 2050, about 30 terawatt-hours of spent batteries from EVs and plug-in hybrid EVs could reach end of life. 
\cite{Xu2023} While recycling can meet 20-30\% of projected Li, Ni, and Co demands by 2050, there remain critical scale-up challenges, specifically how to maintain efficient recovery of products from mixed feedstock while reducing environmental and social impacts.
\cite{iea}

The main methods to recycle Li-ion batteries use high-temperature smelting (pyrometallurgy), aqueous solutions for extraction and recovery (hydrometallurgy), cathode reconstitution (direct recycling), or a combination of these. For example, Umicore Valeas and Glencore use pyrometallurgy to isolate cobalt and hydrometallurgy to recover a flexible range of metal salts. \cite{baum2022} Hydrometallurgy is an advantageous final step in the purification and processing of mixed waste streams, but it often consumes non-regenerated reagents, creates substantial solid waste and highly saline wastewater, and contributes significantly to the carbon footprint. To improve its environmental impact, this inherently ``linear" process must be redesigned according to circular principles to improve environmental impact. \cite{Binnemans2023} Recent efforts have addressed these challenges by focusing on green hydrometallurgical pathways, such as utilizing waste feedstocks, adopting greener organic solvents, and developing methods that minimize acid consumption and energy usage.\cite{ooi2024,huang2025} However, the continued reliance on mineral acids remains problematic due to their wastefulness, corrosivity, and disposal challenges, while the use of even greener organic solvents still faces issues such as toxicity, limited biodegradability, flammability, and difficulty in regeneration. 

On this end, green designer solvents consisting of environmentally-benign components, such as inexpensive type-III deep eutectic solvents (DESs), may have advantages as they reduce operating temperature, reaction time, and have lower toxicity.\cite{Padwal2022} It was first shown by Tran \textit{et al} \cite{Tran2019} that the ethaline DES, consisting of a 2:1 molar ratio of ethylene glycol (EtGl): choline chloride (ChCl), can leach Co and Li with more than 90\% efficiency at 180 \textdegree C. Yet, at this temperature, ethaline decomposes into toxic and hazardous byproducts, trimethylammine and 2-chloroethanol, limiting its green advantage. \cite{Peeters2022} Nevertheless, for DESs to reach relevance in industrial applications, it is their long-term thermal stability which should be practically assessed. \cite{YChen2021,Marchel2022}

In this work, we evaluate the long-term thermal stability of ethaline at 60 \textdegree C, and report that ethaline in fact exist as a partially-decomposed solvent containing ChCl, EtGl, toxic chloromethane (\ce{CH3Cl}), and dimethylaminoethanol (DMAE). To study these solvent-assisted decomposition mechanisms with quantum mechanical (QM) accuracy, we use state-of-the-art machine learning interatomic potentials (MLIPs). We reveal that the dynamic hydrogen (H)-bond network unlocks a metastable potential energy surface within ethaline, lending to unfavorable configurations such as Cl trapping near electrophilic sites, initiating decomposition via \ce{S_N 2} reaction. Evidently, future design of DESs should consider the strength of the hydrogen bond donor-acceptor pair to avoid self-reactivity and decomposition in hydrogen-bound mixtures.



\begin{figure}
    \centering
    \includegraphics[width=0.4\textwidth, center]{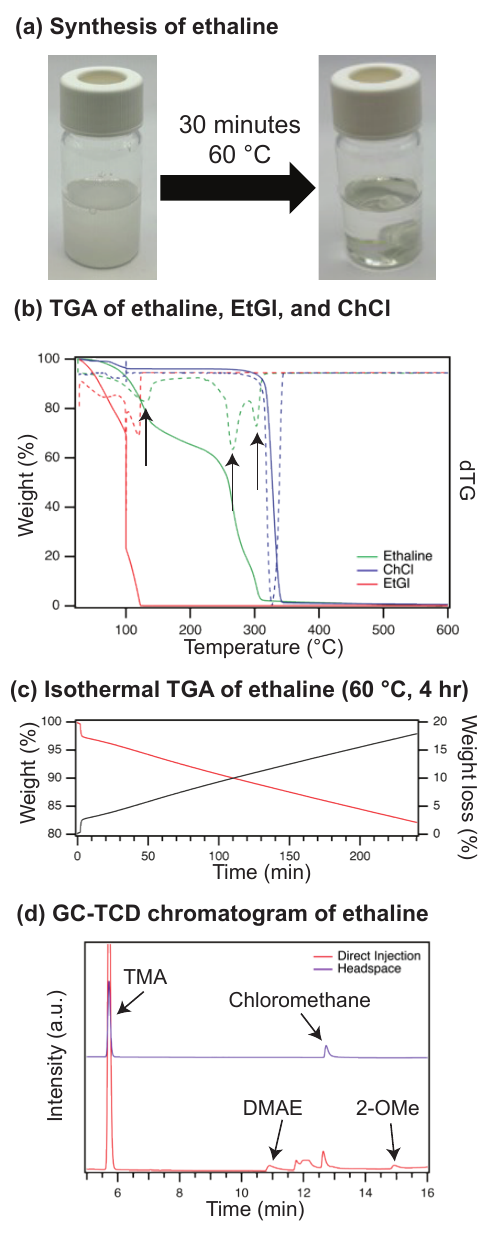}
    \caption{Synthesis and characterization of ethaline. (a) Images of ethaline at t = 0 min. and after 30 minutes (b) Dynamic thermogravimetric analysis (TGA) of ethaline (solid), ethylene glycol (EtGl), and choline chloride (ChCl). Dashed lines are dTG, with inflexions noted (arrows). (Ramp rate: 5 \textdegree C/min, 25 – 600 \textdegree C) (c)  Isothermal TGA of ethaline (60 °C for 4 hours) and (d) Representative GC-TCD chromatogram showing the composition of the ethaline (red) and vial headspace (blue) sample obtained after 30 minutes.
}
    \label{fig:experiments}
\end{figure}

The stability of ethaline (EtGl:ChCl in a 2:1 molar ratio) is experimentally investigated at 60 \textdegree C (below 80 \textdegree C, the typical synthesis condition for ethaline). The mixing of EtGl (clear liquid) and dried ChCl (white flakes) resulted in a white slurry that became colorless after stirring for 30 minutes when heated at 60 \textdegree C, achieving a viscous consistency (Fig. \ref{fig:experiments}a). The water content of ethaline after 30 minutes of mixing was around 0.77 wt.\%, which is attributed to the hygroscopic nature of the system (Table \ref{stable:water-content}). This level of water content aligns with typical conditions reported in the literature and reflects practical preparation conditions. \cite{wu2023,Valverde2020}

Dynamic thermogravimetric analysis (TGA) with a heating rate of 10 K/min on the prepared ethaline and its pure constituents is shown in Fig. \ref{fig:experiments}b. The DES demonstrates higher thermal stability than pure EtGl but lower stability than ChCl, largely due to the volatility and thermal instability of EtGl before its boiling point, a characteristic commonly observed in DESs. \cite{rodrigeuz2018} At the same time, inflections in the derivative of the thermogram (dTG) at 131, 266, and 302 \textdegree C (black arrows, \ref{fig:experiments}b) indicate the presence of pre-existing decomposition products formed during preparation. As dynamic TGA tends to overestimate thermal stabilities, isothermal TGA was carried out at a constant temperature of 60 \textdegree C, showing a significant mass loss of 17 wt.\% after 4 hours (Fig. \ref{fig:experiments}c). 

Mass loss due to decomposition was further characterized using gas chromatography (GC), following previously reported methods. \cite{rodrigeuz2018, Peeters2022} A direct injection of a 5 $\mu L$ liquid sample of ethaline, along with headspace injection (50 $\mu L$) from the vial, revealed peaks at 5.5 and 12.6 minutes (Fig. \ref{fig:experiments}d), indicating the presence of volatile gases such as trimethylamine (TMA) and chloromethane (Fig. \ref{sfig:gc-tcd}), which are decomposition products of ethaline. The liquid sample also showed additional peaks at 10.9 and 14.8 minutes, corresponding to decomposition products such as dimethylaminoethanol (DMAE) and 2-methoxyethanol (2-OMe). Specifically, DMAE was identified at the 11-minute peak and quantified at approximately 26 mM (Fig. \ref{sfig:dmae-conc}). These observations of long-term ethaline decomposition at moderate temperatures are consistent with previous work.\cite{YChen2021}

The thermal stability of deep eutectic solvents (DESs) has been increasingly studied in recent years, focusing on the impact of preparation temperatures on their stability. While instability due to the hydrogen bond donor (HBD) has been noted in a few studies,\cite{rodrigeuz2018,Bennett2005SAXsDP}, this work, along with others, indicate that the hydrogen bond acceptor (HBA) may play an equally or even more significant role in driving decomposition.\cite{Peeters2022,vandenBruinhorst2023} ChCl undergoes a well-known solid-solid transition at 79 \textdegree C, with the $\alpha$-phase exhibiting increased susceptibility to hydration-induced instability. \cite{Marchel2022,ferreira2022} This sensitivity likely stems from its structural configuration and interaction with water. Interestingly, decomposition products were detected regardless of whether ethaline was prepared in the $\alpha$-phase or $\beta$-phase regions, suggesting a more intricate interplay between HBD and HBA decomposition mechanisms. Thus, we employ molecular modeling to explore how decomposition pathways arise from intermolecular interactions and provide predictive insights for the design of new, stable solvent mixtures. 

\begin{figure}
    \centering
    \includegraphics[width=\linewidth]{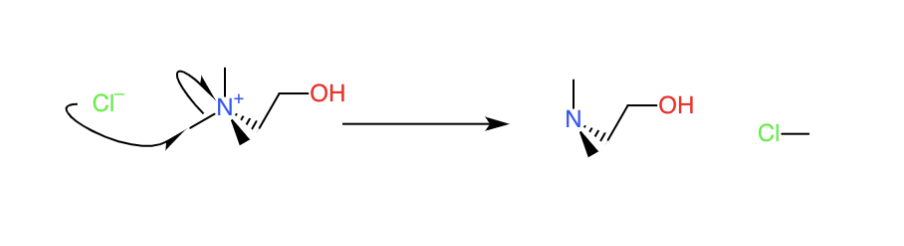}
    \caption{Choline decomposition via \ce{S_N 2} reaction, with chloride and choline as reactants (left) and dimethylaminoethanol (DMAE) and chloromethane \ce{CH3Cl} as products (right).}
    \label{fig:sn2}
\end{figure}

The decomposition of ethaline into DMAE and \ce{CH3Cl} can occur via the \ce{S_N 2} reaction \cite{Datta2023} (Fig. \ref{fig:sn2}), where nucleophilic \ce{Cl} approaches electrophilic \ce{C} from the \ce{CH3} group via backside attack, i.e. the vector connecting \ce{N} and \ce{C}. We define $\chi$ as the collective variable that describes this reaction as: $\chi = r_{\ce{Cl}-\ce{C}}-r_{\ce{C}-\ce{N}}$, where $r_{\ce{Cl-C}}$ is the distance between nucelophile \ce{Cl} and electrophile \ce{C}, and $r_{\ce{C-N}}$ is the distance between the electrophile \ce{C} and leaving group (\ce{N} of DMAE). \cite{tutorial-us} This collective variable is convenient for describing reactants ($\chi>0$), products ($\chi<0$) and the transition state ($\chi\approx0$).

\begin{figure}
    \centering
    \includegraphics[width=0.5\linewidth]{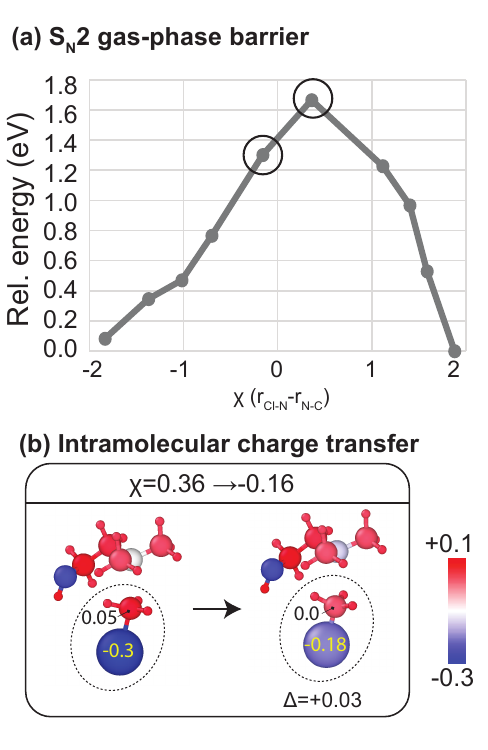}
    \caption{\ce{S_N 2} decomposition of ChCl in gas-phase. (a) Reaction pathway along the collective variable, $\chi$, of the \ce{S_N 2} decomposition of choline chloride, calculated using CI-NEB and DLPNO-CCSD(T). Reactant and product states correspond to $\chi>0$ and $\chi<0$, respectively. The most significant intramolecular charge transfer occurs when $\chi\approx 0$  (circled). (b) Intramolecular charge transfer (computed by Hirshfeld charges) near the transition state ($\chi=0.36\rightarrow\chi=-0.16$) occurs in two steps, where nucleophilic \ce{Cl} is further oxidized and electrophilic \ce{C} and \ce{H} are further reduced. \ce{CH_3Cl} is overall oxidized by  $\Delta=+0.03$ (indicating that all of DMAE is reduced by 0.03). The charge-transfer is instantaneous (black arrow) in a non-rate-limiting environment.}
    \label{fig:gas-phase-barrier}
\end{figure}

Using DLPNO-CCSD(T) \cite{dlnpo-2013a,dlpno-2013b}, we calculate that the gas-phase barrier (Fig. \ref{fig:gas-phase-barrier}) is 1.68 eV, which is higher than the $\approx$ 0.6 eV barrier for the gas-phase \ce{[Cl(CH_3)Cl]-} \ce{S_N 2} reaction \cite{Chandrasekhar1984}, but aligned with understandings that increasing bulkiness of the substrate (from \ce{CH_3} to \ce{(CH_3)_3N-(CH_2)}, in this case) can significantly increase the activation barrier. \cite{gallegos2022} The products are destabilized by 82 meV in vacuum. 

In solution, the reaction barrier is expected to increase given that the products are non-ionic in a polar environment. \cite{Chandrasekhar1984} The extent of this increase can be probed using molecular modeling. Furthermore, in the ethaline solvent, intramolecular charge-transfer is rate-limited by solvent reorganization due to sluggish rotation of choline occurring over 200 ps characterized using femtosecond time-resolved absorption spectroscopy. \cite{Alfurayj2021} This suggests that the charge transfer near the transition state ($\chi\approx0$) is also expected to occur over 200 ps in ethaline (Fig. \ref{fig:gas-phase-barrier}b). 

This time scale is beyond \textit{ab initio} molecular dynamics (AIMD) and different approaches are needed. Recently, the free energy of peptide bond formation in explicit water was calculated via umbrella sampling by Rolf \textit{et al} using MLIP.\cite{rolf2024} The authors trained a DeepPMD potential on more than 76,000 configurations including biased trajectories, propagated with enhanced sampling (steered MD, metadynamics, and umbrella sampling).  However, generating  a large number of configurations to achieve this stability is computationally challenging, particularly when using hybrid functionals. Furthermore, as solvation environments increase in complexity, which is the case with ethaline compared to water, it is likely that more umbrella sampling configurations are needed to train stable and representative MLIPs capable of umbrella sampling. We show that another method, described below, can serve as an approximation of the dynamic reaction mechanism in solution, where the effect of explicit solvation can be taken into account.

\begin{figure}
    \centering
    \includegraphics[width=0.40\textwidth, center]{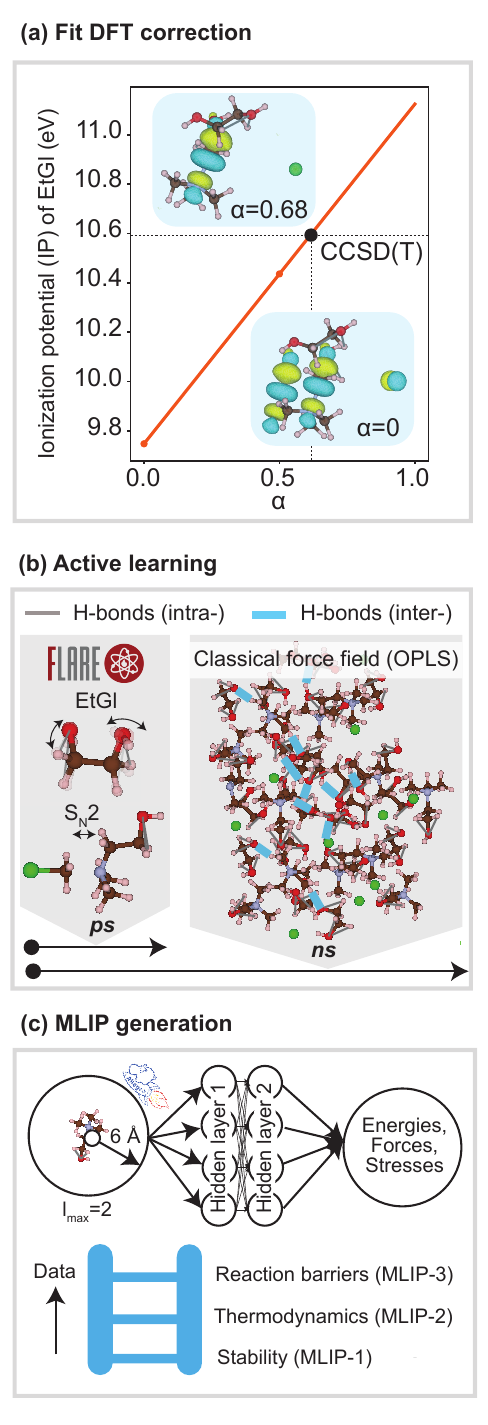}
    \caption{Computational workflow to study ethaline. (a) 0.68 HF exchange removes artificial charge transfer in ethaline, evidenced by the LUMO on Cl for $\alpha=0$, and no LUMO on Cl for $\alpha=0.68$. Yellow and cyan coloring are (+) and (-) isovalues. (b) Active learning from the FLARE package captures intramolecular H-bonds (grey lines), while OPLS captures intermolecular H-bonds (blue thick lines). The \ce{S_N 2} pathway is also sampled by FLARE. (c) The MLIP, with key parameters noted, exhibits model stability, but undergoes additional training to attain correct thermodynamics (second rung) and reaction barriers (third rung).}
    \label{fig:computational-workflow}
\end{figure}

Prior works studying chemical reactivity in various environments have used active learning with QM calculations to build machine learning interatomic potentials (MLIPs). \cite{Yang2022,Young2021,Zhang2024} Fig. \ref{fig:computational-workflow} summarizes our developed workflow, also using active learning to build a MLIP, albeit starting from establishing the DFT approximation. Following previous procedures \cite{Park2017,Falletta2020}, we use PBE0 (25\% Hartree Fock (HF) exchange) \cite{Adamo1999} and tune the HF correction by fitting to the gold-standard reference, CCSD(T), for the ionization potential (IP) of EtGl(g). The resultant correction is $0.6851$ HF exchange, which we call PBE0(68)-D3. All details of DFT settings using CP2K \cite{cp2k} and Orca \cite{Neese2020} and verifications with CCSD(T) and DLPNO-CCSD(T) are in the Supporting Information (SI). 

We use active learning \cite{Vandermause2022}, classical force fields\cite{Jorgensen1996}, and iterative training to construct our equivariant neural network MLIP \cite{Musaelian2023}. Active learning expedites the sampling of intramolecular degrees of freedom occurring over picoseconds (ps) time scale (e.g., intramolecular H-bonding), while classical force fields capture intermolecular degrees of freedom (e.g., intermolecular H-bonding) occurring over nanosecond (ns) timescale. Fig. \ref{sfig:flare-opls-bl} describes the bond length diversity sampled from active learning. 

\begin{figure}
    \centering
    \includegraphics[width=0.40\textwidth, center]{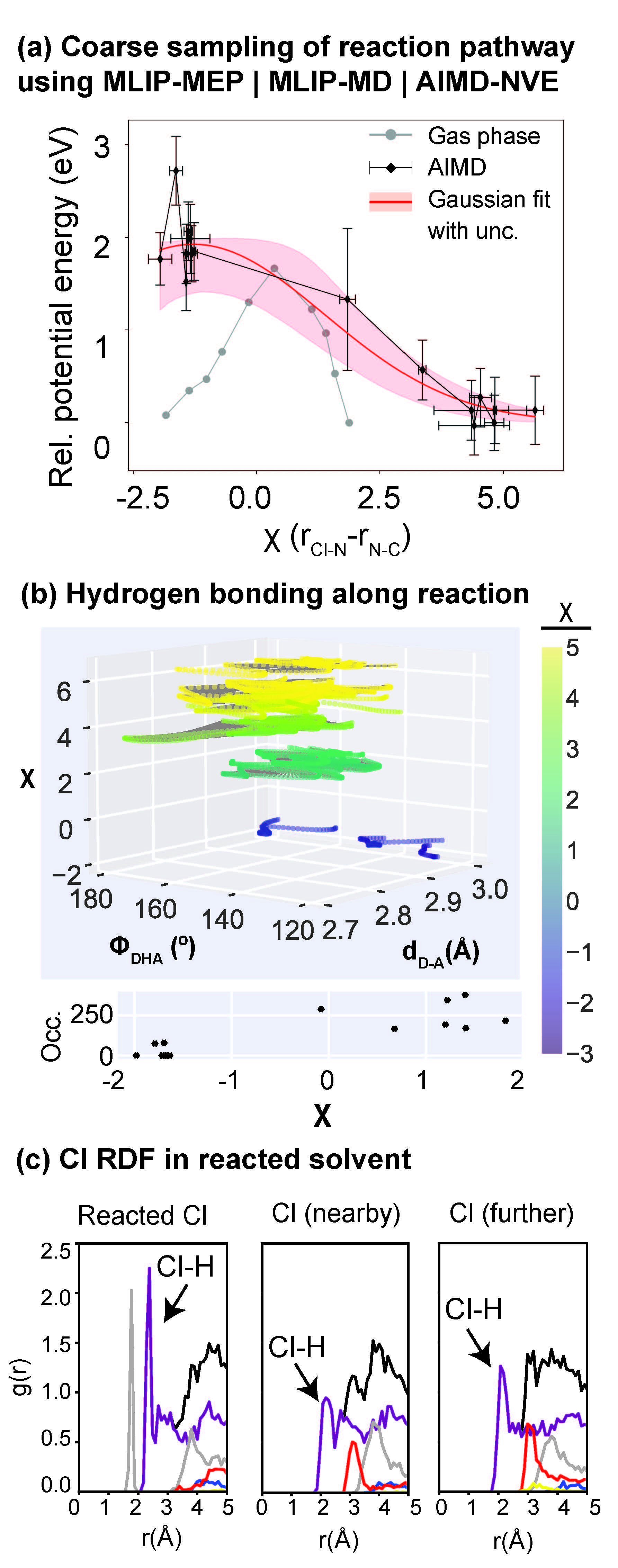}
    \caption{\ce{S_N 2} decomposition in ethaline solvent.(a) Results of the coarse-sampling along $\chi=r_{\ce{Cl-N}}-r_{\ce{N-C}}$, using MEP calculations, followed by 200 ps of MLIP equilibration and 500 fs of NVE AIMD (shorthand: MLIP-MEP |MLIP-MD | AIMD-NVE). Energies from AIMD (mean and standard deviation) are in black, the Gaussian($+1\sigma)$ function fit is in red, and vacuum \ce{S_N 2} barrier calculation, juxtaposed for context, is in grey.  
    (b) H-bonding is characterized by the bond angle between \ce{O-H-Cl} (donor-H-acceptor), $\phi_{DHA}$ and distance between donor and acceptor, $r_{D-A}$. Grey lines connect H-bonds which appear in the same frame. The occurrence (Occ.) of H-bonds is the maximum continuous accumulation of H-bonds around the reacting \ce{Cl} during NVE AIMD. (c) The RDF in reacted solvent, around various solutes (reacted Cl, nearby Cl, and further-away Cl) from $r=0$ to $5$ \AA, showing Cl-H (purple), Cl-C (grey), Cl-O (red), Cl-N (blue), Cl-Cl(yellow), and Cl-all (black).}
    \label{fig:barrier}
\end{figure}

Iterative training then patches the failure modes of MLIP, which are unphysical bond breaking (Fig. \ref{sfig:allegro-0-4ps}) and over-stabilization of reaction intermediates (Fig. \ref{sfig:baseline-mep}). However, when retrained, the final MLIP reproduces the potential energy surface (PES) and bulk structure at 298 K (Fig. \ref{sfig:nvt-neat-bulk-model},\ref{sfig:nvt-rdf-comparison}) and \ce{S_N 2} reaction pathways (Fig. \ref{fig:barrier} and Fig. \ref{sfig:sn2-mep}). Iterative training has been observed to be an important step in generating physically-reasonable potentials. \cite{Magdau2023,Goodwin2024} 

We use MLIP-3 from the third ladder rung (Fig. \ref{fig:computational-workflow}c) to sample configurations of ethaline in NVT at 25 \textdegree C, construct Minimum Energy Pathways (MEP) of the \ce{S_N 2} reaction \cite{Henkelman2000A}, and simulate solvent relaxation occurring over 200 ps.  \cite{Alfurayj2021} (The computed self-diffusivity is consistent with experiments, enabling comparison (Fig. \ref{sfig:msd}).) No solvation reorganization occurs during the MEP calculation, as all forces are optimized, but the solvent is effectively ``rigid" as rotational, translational, or vibrational modes are not sampled. 

After 200 ps of equilibration, each frame is deployed for $\approx$ 500 fs in AIMD NVE using PBE(0)68-D3, with average simulation temperatures of 284-292 K (Table \ref{stable:solvent-relax-temp}). The purpose of the NVE production trajectory is to characterize the dynamic H-bond contributions to the PES along the reaction pathway. 

Results of this coarse sampling approach are shown in Fig. \ref{fig:barrier}a. Since the equilibration is dynamic, there are deviations in $\chi$ (horizontal error bars) and in the relative PES (vertical error bars). Overall, the reaction is uphill, and products are destabilized by 2 eV (c.f. vacuum: 82 meV), confirming expectations that non-ionic molecules are de-stabilized in polar environments. The predicted reaction energy, fitted by the Gaussian with $1\sigma$ uncertainties shown, is also higher than in vacuum. Near the transition state ($\chi=1.72$), there is a large, 700 meV deviation in the PES. We distill all deviations in $\chi$ and the relative PES in terms of the local H-bonding environment around the reacting \ce{Cl}.  

Fig. \ref{fig:barrier}b quantifies the presence of H-bonds around the reacting \ce{Cl} in terms of the donor-H-acceptor bond angle, $\phi_{DHA}$, and donor-acceptor distance, $r_{DA}$, as a function of $\chi$. An H-bond is considered formed if  $120^{\circ} < \phi_{DHA}<180^{\circ}$ and $r_{DA}< 3$ \AA. Additionally, the ``persistence" of H-bond(s) around the reacting \ce{Cl} is evaluated by counting the maximum, continuous, cumulative occurrence of H-bonds over $\approx$ 500 fs: one H-bond with reacting \ce{Cl} in a frame is one occurrence; two H-bonds with reacting \ce{Cl} in a frame is two; no H-bonds to reacting \ce{Cl} resets occurrences to 0. 

Before the reaction ($\chi\approx5$), Fig. \ref{fig:barrier}a-b shows how the persistence of H-bonding with \ce{Cl}, spanning a range of angles and distances, perturbs the PES by $\pm$ 300 meV and enables relatively mobile \ce{Cl}. Closer to the transition state($\chi=1.72$), this positional flexibility of \ce{Cl} is traded for metastability. Here, the H-bonding now perturbs the PES by $\pm$ 700 meV and \ce{Cl} is firmly locked in place, experiencing a narrower range of H-bonding, in terms of $\chi$, $\phi_{DHA}$, and $r_{DA}$. 

Strong and very strong H bonds are generally found in systems with cations or anions, which is the case for ethaline. Very strong H-bonds (> 1 eV), in this case, are not formed. \cite{HIBBERT1990255} We surmise that formation of the strong H-bonds which lock \ce{Cl} in place in high-energy states near reaction sites may be especially prominent in ethaline as both EtGl and Ch contain H-bond donor groups (\ce{OH}). In ethaline, the 2:1 molar ratio of EtGl:ChCl means that for every one \ce{Cl}, there are five \ce{OH} groups. These prevalent, persistent interactions with \ce{Cl} could be among the reasons that the computed self-diffusivity of \ce{Cl} is even less than choline (Fig. \ref{sfig:msd}) despite it being smaller and more symmetric. At moderate temperatures of 60 \textdegree C, the \ce{S_N 2} reaction is expected to occur more frequently. Once a \ce{Cl-} is locked into place by H-bonds, it is not easily broken by thermal fluctuations ($\approx$ 28 meV at 60 \textdegree C). 




Fig. \ref{fig:barrier}c shows how, after the reaction, a local void forms around \ce{CH3Cl}, and \ce{Cl-H} intermolecular bonding no longer exists. The formation of \ce{CH3Cl} permanently weakens the H-bond network for nearby \ce{Cl-} compared to \ce{Cl-} further away. The formation of neutral DMAE and \ce{CH3Cl} with reduced H-bonding is consistent with the understanding that neutral systems typically exhibit weak H-bonding. \cite{HIBBERT1990255}

Future design of green solvents may benefit from detailed explicit analysis of hydrogen bond donor-acceptor interactions along various decomposition pathways, which is captured in Fig. \ref{fig:barrier}. We have shown that a coarse approximation to the reaction pathway, namely using MLIP-MEP | MLIP-MD | AIMD-NVE simulations to relax solvation degrees of freedom beyond the computational reach of hybrid-DFT AIMD can generate molecular-level resolution of reaction pathways. Although the approximation by no means circumvents the utility of umbrella sampling, it may enable greater throughput, albeit lower-resolution, evaluation of new designer solvents because significantly fewer training data are required (Table \ref{stable:allegro-errors}). Novel solvents can be vetted for stability and reactivity following a similar approach. 

As a last point, we note that during the MLIP-MEP calculation, the MLIP preferentially moves MEP images away from the transition state (Fig.  \ref{sfig:mlip-relaxation-trj}); thus only one image near the transition point, $\chi=1.72$, is captured. More details and discussions are in the SI. 

In conclusion, ethaline, a 2:1 molar ratio of choline chloride and ethylene glycol, has been evaluated for long-term thermal stability. We experimentally find evidence of decomposition of neat solvent into dimethylethanolamine, chloromethane, trimethylamine, and 2-chloroethanol at 60 \textdegree C and study the reaction mechanism via \ce{S_N 2} decomposition. The origin of the reaction arises from H-bond formation which trap \ce{Cl-} near the reacting site. Whether this behavior is primarily due to the choice of ethylene glycol as a hydrogen bond donor remains to be fully explored; however, it is worth noting that other choline chloride-based deep eutectic solvents also demonstrate poor long-term thermal stability. \cite{Marchel2022} This work motivates studies exploring green designer solvents to 
prioritize the hydrogen bonding strength as key selection criterion for proposing new stable solvent-mixtures.



\section*{Experimental Methods}
\subsection*{DFT calculations}
All of the energy and force evaluations from density functional theory (DFT) calculations are carried out using the QUICKSTEP module of the CP2K package (version 2023.1.). \cite{cp2k} The molecular triple-$\xi$ basis set (TZVP-MOLOPT-GTH), auxiliary density matrix method with basis set pFIT3, and the GTH-PBE pseudopotentials are used for all atoms. No purification method for wavefunction fitting is used and EPS\_SCF is set to 1.0 E-6. We use the generalized gradient approximation (GGA) Perdew-Burke-Erzenhof (PBE) functional \cite{pbe} with varying \% (referred to as $\alpha$) of Hartree-Fock (HF) exchange to fit the ionization potential (IP) of gas-phase ethylene glycol (EtGl) at the coupled cluster single double (triple) (CCSD(T)) level with the augmented correlation-consistent triple-$\xi$ basis set (aug-cc-pVTZ), using the default integration grid (defgrid2) and SCF convergence tolerance (NormalSCF) in ORCA. \cite{Neese2020} A cutoff of 500 Ry (CUTOFF) and a relative multi-grid cutoff of 50 Ry (REL\_CUTOFF) are used for all calculations, and convergence is checked for both (for CUTOFF: the total energy changes less than 0.1 meV/atom relative to the cutoff at 1000 Ry (Fig. \ref{sfig:convergence-test}); for REL\_CUTOFF: the total energy changes less than 0.01 meV/atom after a relative cutoff of 30 Ry is used(Fig. \ref{sfig:convergence-test-relative-cutoff}). A cutoff radius of 5 \AA\ is used for the truncated Coulomb interaction potential, and shown in a convergence test for a cell size of 13 $\times$ 13 $\times$ 12 \AA (Fig \ref{sfig:truncation-radial-convergence}). A sample input script for all CP2K calculations is provided in Supporting Information. 

\subsection*{Classical Force Fields}

To obtain structures with intermolecular diversity spanning ns of simulation time, we utilise classical force fields to sample representative structures. We generate a box with 6 ChCl and 12 EtGl with dimension of 15.85 \AA\ using packmol and fftool. We utilise the CL\&P force field for ChCl, and the OPLS-AA force field for EtGl. A timestep of 0.5 fs was utilised with the velocity-verlet to evolve the equations of motion, the Nos\'e-Hoover barostat (with the coupling set to 1000$\times$ the timestep) and thermostat (with the coupling set to 100$\times$ the timestep). Several temperatures are investigated: 300~K, 400~K, 500~K and 1000~K. All structures are initially equilibrated for 10~ns at their respective temperatures in NPT, before a NPT production run where structures are saved every 10~ps.

Although the non-polarizable OPLS force field developed by Doherty and Acevedo \cite{Doherty2018} for ethaline could have also been used to sample intermolecular diversity, we use the more general OPLS-AA force field to enable extendable workflow to other solvents that may not have refined force fields. Additionally, the classical force field is a coarse sampling method complementing the intra-molecular diversity sampled through active learning, creating a more diverse dataset to train on. 

\subsection*{Active learning}

To sample neat solvent configurations using active learning, we geometrically relax isolated molecules of choline (Ch) and EtGl, in PBE(0)68-D3. Then, molecules are placed randomly in a box using the Packmol,\cite{packmol2009} ranging from system sizes of 168 atoms (4 Ch, 4 Cl, and 8 EtGl molecules; box size 13 \AA\ by 11.81 \AA\ by 11.99 \AA), 210 atoms (5 Ch, 5 Cl, and 10 EtGl molecules; box size 14.9 \AA\ by 12.6 \AA\ by 12.36 \AA), and 252 atoms (6 Ch, 6 Cl, and 12 EtGl molecules; box size 13.53 \AA\ by 14.7 \AA\ by 14.48 \AA). Note that sampling these compositions, with the same ratio of ChCl and EtGl, will allow the models to be energy extensive for this composition, but transferring the model to other compositions will incur energy errors, although as ``high entropy'' compositions are sampled the forces should be transferable. \cite{Goodwin2024} These configurations are then geometrically relaxed again in CP2K-2023.1 using the QUICKSTEP module with our PBE(0)68-D3\cite{cp2k}. 

The final configurations are used as the initial structures for an active learning workflow using the FLARE code, with the Velocity-Verlet to evolve the equations of motion, a timestep of 0.5 fs, training hyperparameters from the second frame onwards. The active learning workflow uses model predicted uncertainty to decide whether to keep exploration configuration space, or call DFT to collect new training data. The uncertainty threshold to call PBE(0)68-D3 is 0.025, and the atoms of uncertainty above 0.0025 are added to the model. Higher thresholds result in broken molecular connections (i.e. fragments), as characterized by SMILES analysis. \cite{pybel} We use the B2 descriptor, $n_{max}=4$, $l_{max}=4$, quadratic cutoff function, $5$ \AA\ cutoff, single neutral atom energies, and all DFT inputs described in Section \ref{SI:dftsetup}. There are 521 frames, ranging from 168 atoms to 252 atoms per frame, collected via FLARE active learning as DFT training data. 

\subsection*{Machine Learning Interatomic Potential training}
We use an optimized version of Allegro\cite{10.1145/3581784.3627041}, with $r_{max}=6 $\AA\ , $l_{max}=2$, and 2 layers. We weight the force, energy, and stress as 1, 100, and 1000 and train on per atom MSE loss, splitting up the training, validation, and test set sizes to be 70\%, 15\%, and 15\%. The learning rate is set to 0.002 and batch size is 2. The strict locality of the potential is not a problem as the cutoff of 6 \AA\ is found for ionic liquids to result in good errors and reasonable density values. \cite{Goodwin2024} Although increasing the cutoff to 7 \AA\ will decrease the errors even further, the low errors in Table \ref{stable:allegro-errors} already reach state-of-the-art. Additionally, all simulations are deployed in NVT, so improvements in density predictions are not critical. 

\subsection*{Synthesis of Ethaline} 

Choline chloride (ChCl, $>98\%$), methanol (MeOH, $>$99.9\%, HPLC grade), ethylene glycol (EtGl,$>$99.8\%), trimethylamine (TMA, 31-35 wt. \% in ethanol, 4.2 M, contains toluene as stabilizer), dichloromethane (1.0 M in diethyl ether), 2-methoxyethanol (2-OMe, 99.8\%, anhydrous) and dimethylaminoethanol (DMAE, $>$99\%) were purchased from Sigma Aldrich. All chemicals were used as received without any further purification. Ultrapure water (18.2 M$\Omega$ cm) was provided by a Millipore Milli-Q system.

Choline chloride (5.294 g, 37.9 mmol, 1.0 molar equivalent) was added to a 20 mL borosilicate vial under ambient conditions. Separately, ethylene glycol (4.708 g, 75.8 mmol, 2.0 eq) was weighed in a syringe and then transferred into the vial. Upon transfer, the formation of bubbles and a noticeable drop in temperature were observed, making the vial feel cold to the touch. The reaction mixture was then heated to 60 \textdegree C and stirred for 30 minutes, resulting in the formation of a transparent, viscous solution. 

\begin{acknowledgement}

 J.H.Y gratefully acknowledges funding from Harvard University Center for the Environment. A.W.S.O. thanks the Shared Materials Characterization Laboratory (Columbia University) for the use of their TGA that was used to analyze ethaline samples. The authors would also like to thank anonymous reviewers for their critical contributions which greatly improved this manuscript. This research was in part supported by the NSF through the Harvard University Materials Research Science and Engineering Center Grant No. DMR-2011754 and
by a Multidisciplinary University Research Initiative sponsored by the Office of Naval Research, under Grant N00014-20-1-2418. Computational resources are provided for by Harvard FAS Research Computing.

\end{acknowledgement}
\begin{suppinfo}

Supporting Information includes: DFT convergence tests, checks on charge delocalization before and after exact exchange correction, verification of PBE(0)68-D3 against DLPNO-CCSD(T), comparisons of PBE(0)68-D3 against R2SCAN charges and NEB, details on active learning trajectories and data collected, further details on the training process for MLIP-0 through MLIP-3 and commentary, radial distribution function (RDF) comparison to previous work, self-diffusivity calculation, MLIP-MEP calculations, equilibration of products and their RDFs, details on the $\approx$ 500 fs AIMD trajectories, water characterization in ethaline, and GC-TCD results. 

All training data, workflow inputs, and MLIPs will be publicly available upon publication. 

\end{suppinfo}

\bibliography{acs-achemso}

\end{document}


\section{Computational Methods}
\label{supplemental:computational}
\subsection{DFT setup}
\label{SI:dftsetup}

To find the amount of exact exchange correction needed for ethaline, we start from the GGA-functional with dispersion corrections, PBE-D3, with tunable HF exchange, $\alpha$

\begin{equation}
    E_{XC} = (1-\alpha)E_X^{PBE} + \alpha E_X^{HF} + E_C^{PBE}
    \label{eq:EXC}
\end{equation}

\begin{figure}
    \centering
\includegraphics[width=\linewidth]{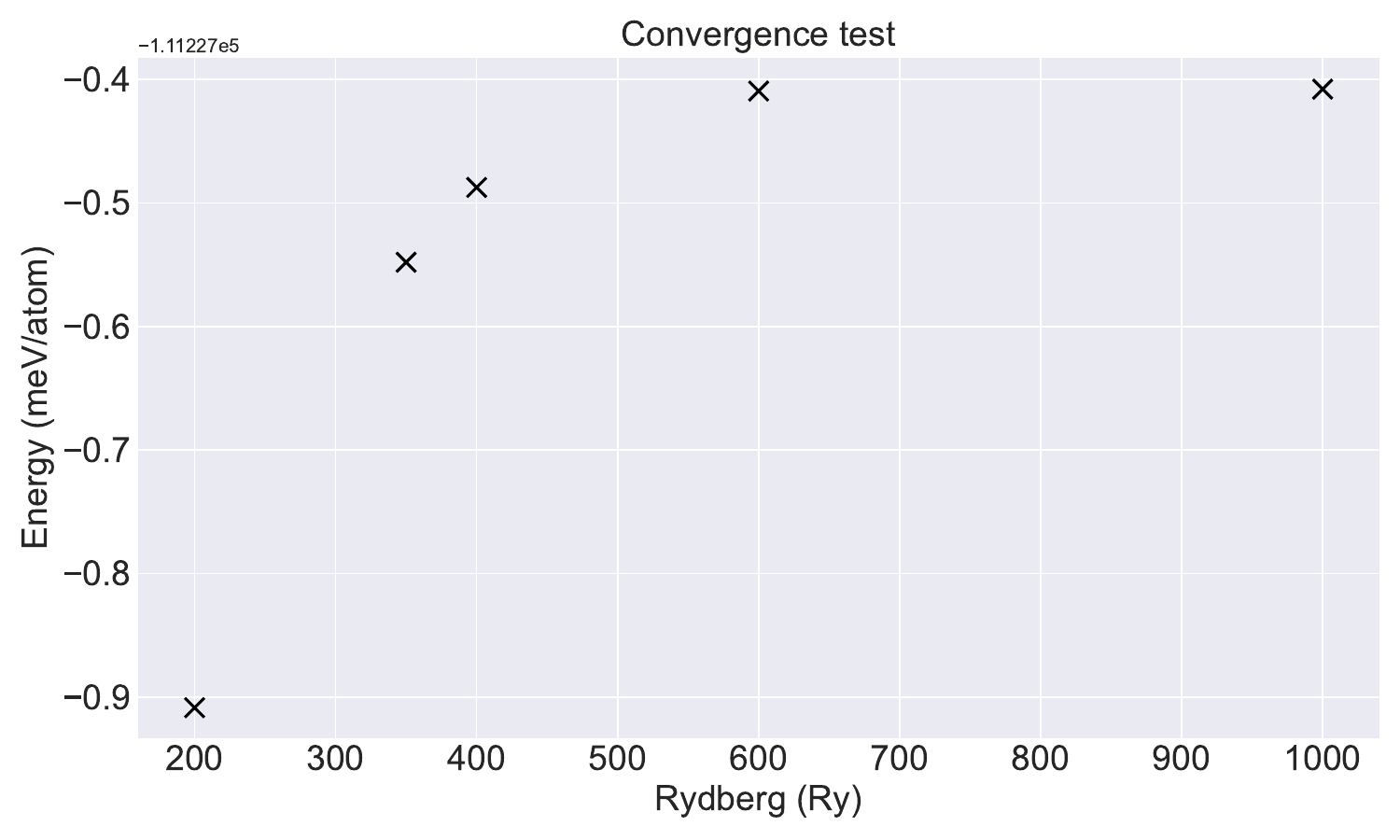}
    \caption{Energy convergence test (CUTOFF) for total energy cutoff for 168 atoms. The final cutoff used for all calculations is 500 Ry or 6804 eV, which is less than 0.1 meV/atom from the energy obtained at 1000 Ry or 13,605 eV.}
    \label{sfig:convergence-test}
\end{figure}

\begin{figure}
    \centering
\includegraphics[width=\linewidth]{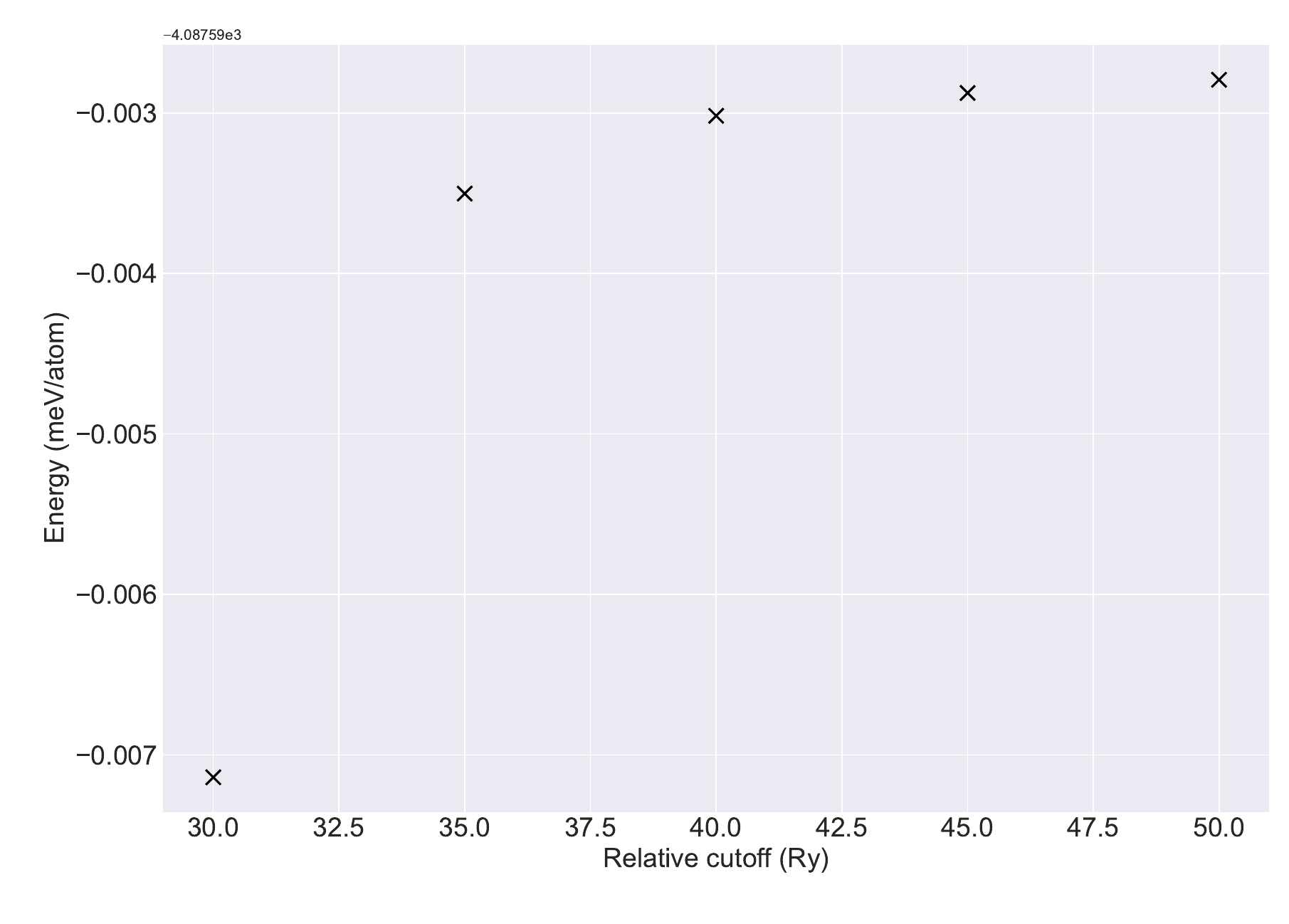}
    \caption{Relative energy convergence test (REL\_CUTOFF) for a total energy cutoff of 500 Ry. Final energies are observed to not vary by more than 0.1 meV/atom after a relative cutoff of 30 Ry is used. The final relative cutoff used for all calculations is 50 Ry.}
    \label{sfig:convergence-test-relative-cutoff}
\end{figure}

\begin{figure}
    \centering
\includegraphics[width=\linewidth]{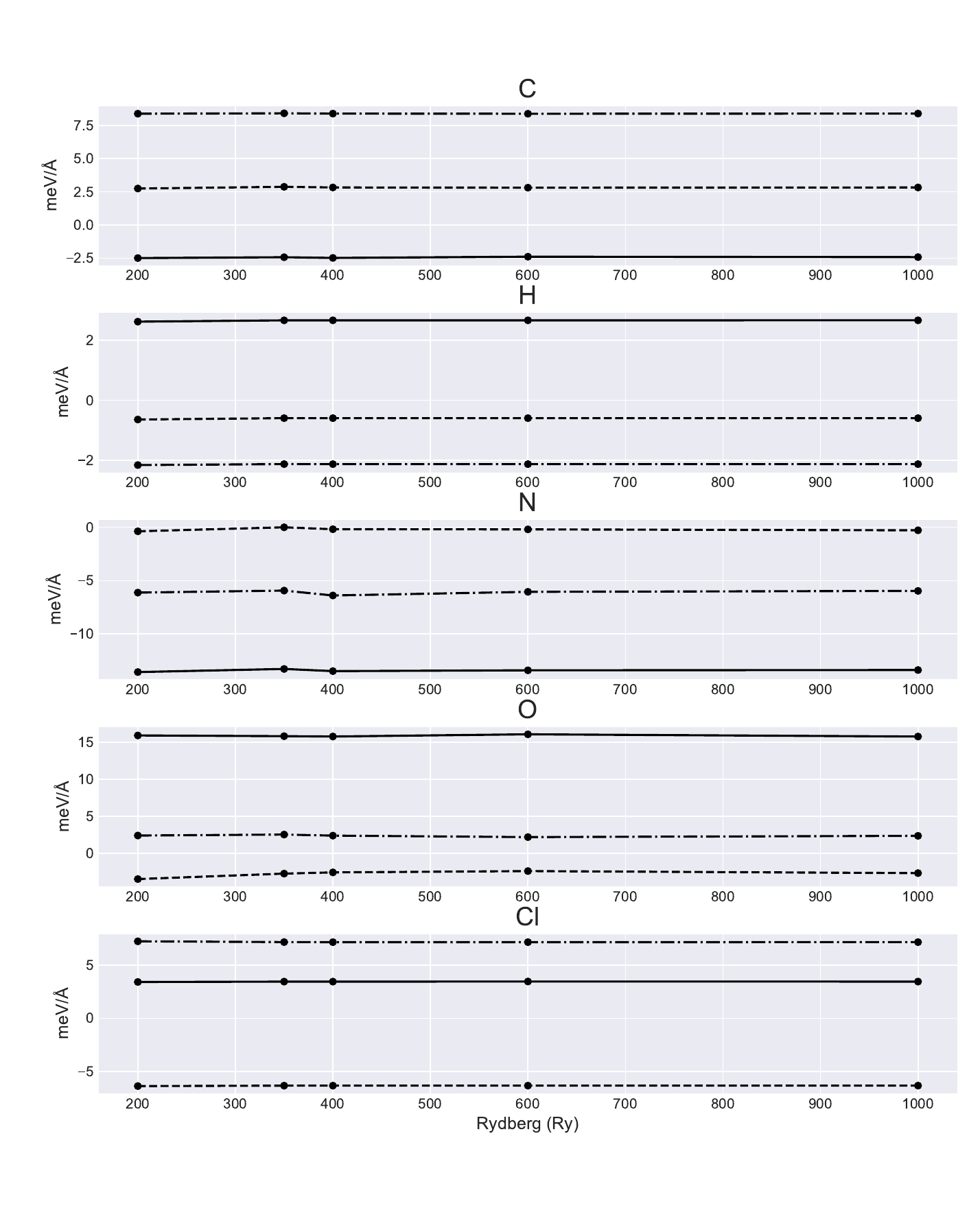}
    \caption{Force convergence test for total energy cutoff for 168 atoms. The final cutoff used for all calculations is 500 Ry or 6804 eV, which shows negligible differences from the forces obtained at a higher cutoff of 1000 Ry or 13,605 eV. Solid, dashed, and dashed-dotted lines correspond to average forces along $F_x,F_y,F_z$ for each specie.}
    \label{sfig:convergence-test-forces}
\end{figure}

Figures \ref{sfig:convergence-test} and \ref{sfig:convergence-test-relative-cutoff} illustrate the convergence with respect to the total energy cutoff and relative energy cutoff in meV/atom. The final cutoffs are 500 Ry and 50 Ry, where total energies change by less than 0.1 meV/atom. Figure \ref{sfig:convergence-test-forces} shows the convergence test for forces (meV/Å) with respect to the total energy cutoff, from 200 Ry to 1000 Ry. The test is shown for each specie (C, H, N, O, and Cl) along each direction ($F_x, F_y, F_z$), with each component averaged across the subset of species in a 168 atom cell. 

\begin{figure}
    \centering 
    \includegraphics[width=0.5\linewidth]{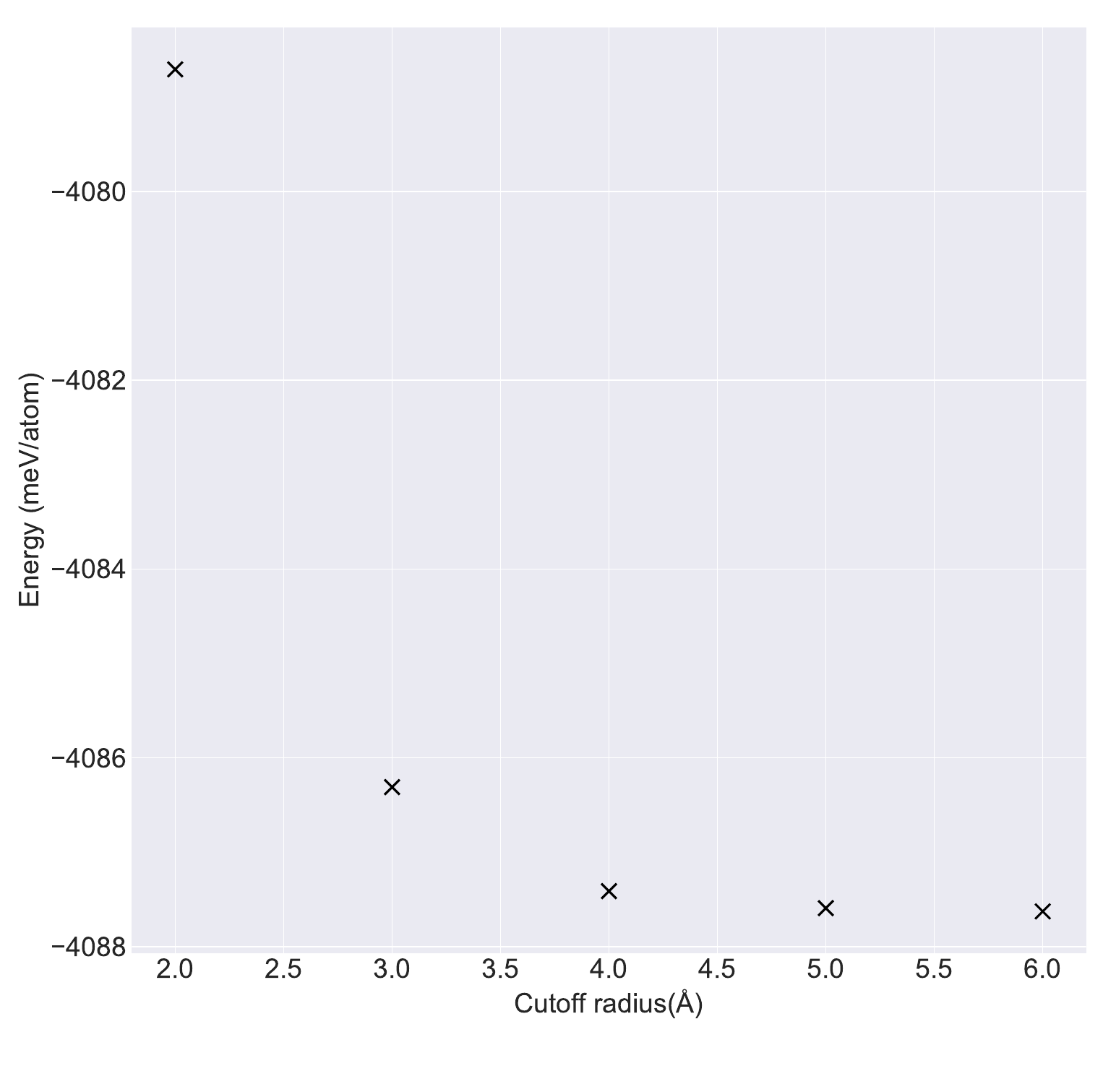}
    \caption{Convergence test for cutoff radius in truncated Coulomb interaction potential. The final cutoff is 5 \AA.}
    \label{sfig:truncation-radial-convergence}
\end{figure}

\clearpage 
Figure \ref{sfig:truncation-radial-convergence} shows the convergence of the cutoff radius for the truncated Coulomb interaction potential in meV/atom. The final cutoff chosen is 5 \AA, and this is observed to not vary by more than 2 meV/atom compared to a longer cutoff of 6 \AA. Note that the cell size of 13 \AA\ by 11.81 \AA\ by 12\AA\ limits the cutoff radius to a maximum of 6 \AA.

\begin{figure}
    \centering
    \includegraphics[width=6.5 in, center]{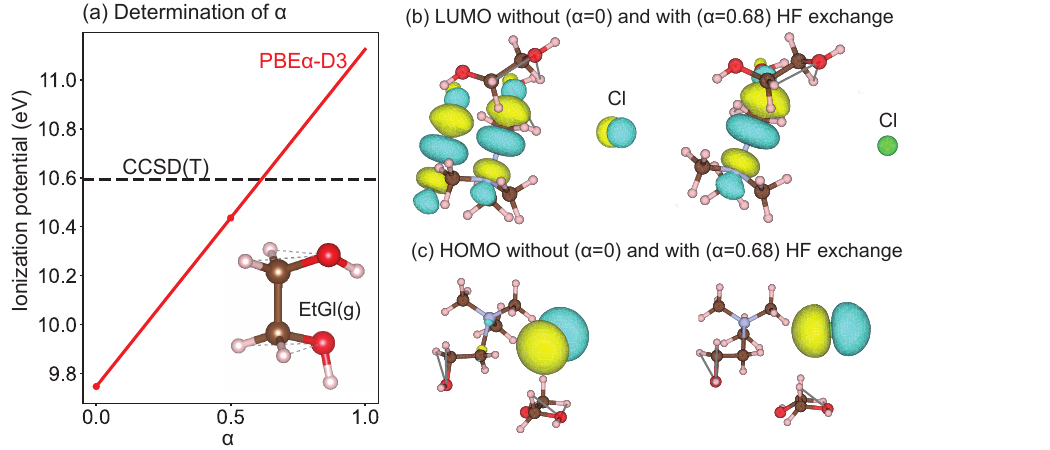}
    \caption{Charge de-localization error in PBE$\alpha$-D3. (a) Determination of $\alpha$ by fitting to IP of gas-phase ethylene glycol (EtGl). (b) LUMO and (c) HOMO of gas-phase choline chloride (ChCl) and EtGl without and with HF correction. Teal and yellow indicate negative and positive charge density isosurfaces, respectively. The isosurface for LUMO is 0.054 and the isosurface for HOMO is 0.031. Red atoms are oxygen, pink are hydrogen, blue is nitrogen, and green is chlorine. Grey dotted lines on molecular cutouts are rendered by Vesta to describe hydrogen-bonding. \cite{Momma2008}}
    \label{sfig:homolumo}
\end{figure}

We use the adiabatic approximation, or $\Delta SCF$ approach, to capture the vertical IP of gas-phase ethylene glycol (EtGl) obtained from NIST \cite{nist-eg-bl}. The results are summarized in Fig. \ref{sfig:homolumo}. This linearity fitting approach has been applied in a range of systems, (e.g. electron donor-acceptor systems and water-splitting). \cite{Park2017,Falletta2020} According to CCSD(T), the IP of EtGl (g) is 10.59 eV, which is reasonably close to the IP found from experiments: 10.21-10.55 eV. \cite{eg-mykyta} Two sets of $\Delta SCF$ calculations with $\alpha=0$ and $\alpha=0.50$ result in a fitted correction of $\alpha=0.6851$, assuming the linearity condition holds. We check that the IP is reproduced by a correction of $\alpha=0.6859$, and find that the IP is 10.49 eV, which is still under-predicted from CCSD(T). However, the amount of exchange is enough to remove the artificial charge transfer observed in PBE-D3. Figs. \ref{sfig:homolumo}b-c shows the HOMO and LUMO of gas phase ethylene glycol before and after the HF correction, confirming that with enough exchange, no spurious oxidation of Cl occurs. 

\subsection{Verification of PBE(0)68-D3 with gas-phase clusters of ethaline}
\label{SI:pbe68-verify}
We take representative cutouts of neat solvent from \textit{ab initio} molecular dynamics (AIMD) simulations and examine their IP in PBE(0)68-D3. For comparison with CCSD(T), the domain-based local pair natural orbital (DLPNO)-CCSD(T) approach with the auxiliary ``/C" basis sets (cc-pVTZ, cc-pVTZ/C) were used with RIJCOSX to speed up the Hartree-Fock step. We use the TightSCF convergence criteria. 

\begin{figure}
    \centering
\includegraphics[width=6 in, center]{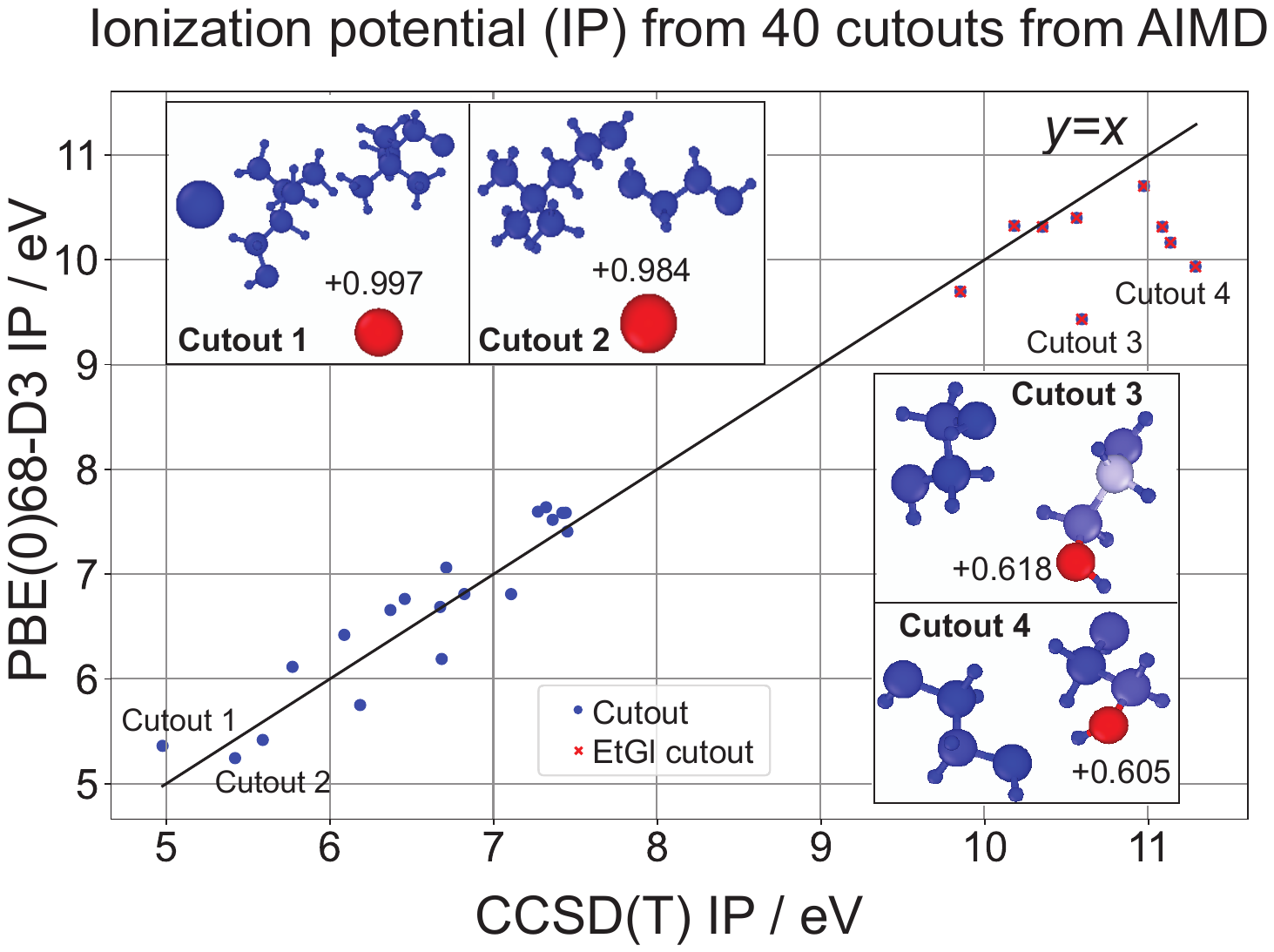}
    \caption{Evaluation of 40 cutouts from AIMD simulations in PBE(0)68-D3 against CCSD(T). Comparison of ionization potentials using the $\Delta$-SCF approach. Red markings indicate of the neat solvent cutouts consist of only EtGl molecules. The four cutouts in the inset show the Hirshfeld charge differences after ionization, where red corresponds to a large positive change and blue corresponds to the least amount of change (close to 0). Cutouts 1 and 2 show that the Cl anion is largely oxidized. Cutouts 3 and 4 show that oxidation is localized to one EtGl molecule. The black line is $y=x$.}
    \label{sfig:ip-cutouts}
\end{figure}

\begin{figure}
    \centering
    \includegraphics[width=6 in, center]{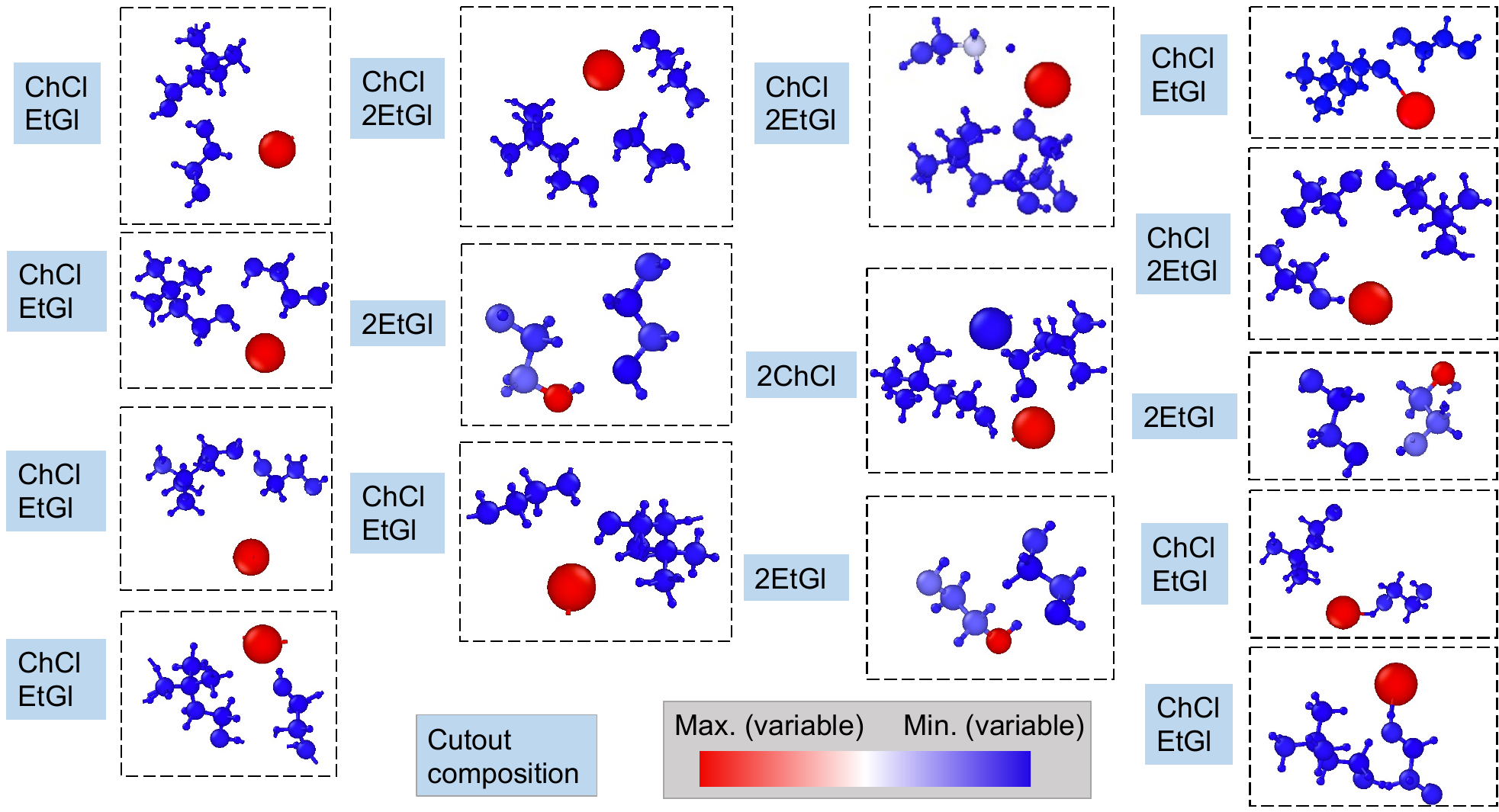}
    \caption{15 cutouts from NVT-AIMD of different compositions: pure EtGl, 2:1 ratio of ChCl:EtGl, and 1:1 ratio of ChCl:EtGl. Each cutout shows the change in Hirshfeld charges, adjusted per frame, to illustrate the extent of the frame-specific charge (de)localization.}
    \label{sfig:morecutouts}
\end{figure}

Starting configurations are initiated using Packmol \cite{packmol2009} in a box containing 6 molecules of EtGl and 3 molecules of ChCl (cell-size: 12.62 \AA\ by 10.62 \AA\ by 15 \AA) . The box is equilibrated for a time of 1 ps, and production run of 2 ps, using the Nos\'e-Hoover thermostat at 298 K with a 0.5 fs timestep. We sample every 100 fs, and collect 40 cutouts of various uncorrelated samples. Fig. \ref{sfig:ip-cutouts} shows the results of IP calculations with respect to DLPNO-CCSD(T), along with the line of best fit, fixing the intercept to (0,0). We find reasonable agreement ($R^2=0.997$), with the exception of some EtGl clusters being outliers which is not unexpected due to the 0.1 eV under-prediction of the IP of EtGl(g). Again, this under-prediction does not translate to charge delocalization errors, as shown in the inset (cutouts 3 and 4). Fig. \ref{sfig:morecutouts} shows the change in Hirshfeld charge during ionization for 15 cutouts (including pure EtGl), illustrating how none have significant charge delocalization error. The charges are normalized per frame to allow for clearer evaluation of extent of charge (de)localization. 

While correction is significantly higher than the 30\% baseline established by Grimme \textit{et al}\cite{Grimme2012}, in organic battery electrolytes, only functionals with 100\% HF exchange reproduce ionization potentials. \cite{Fadel2019}

\begin{figure}
    \centering
    \includegraphics[width=6.5 in, center]{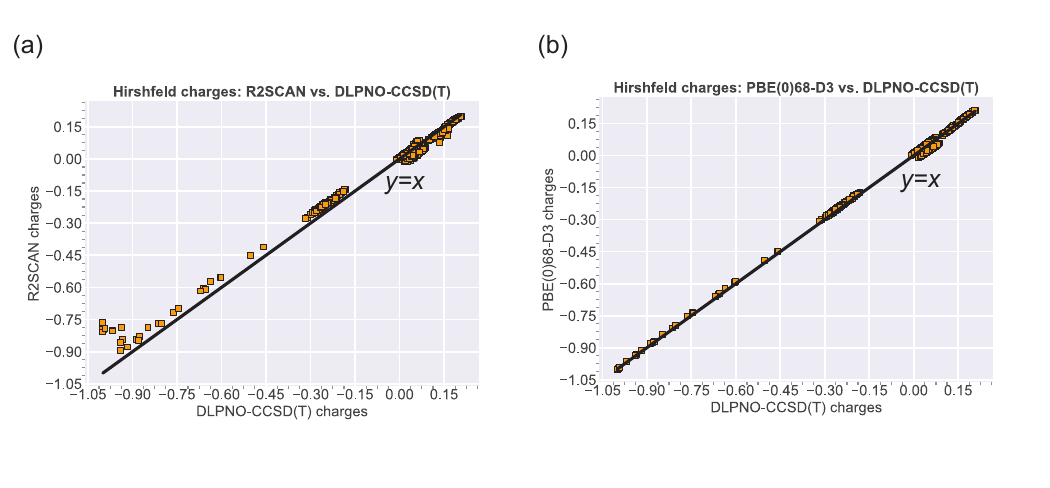}
    \caption{Comparison of Hirshfeld charges on ethaline cutouts computed by (a) R2SCAN and (b) PBE(0)68-D3. The $y=x$ trendline is plotted to highlight systematic error (artificial oxidation) seen in R2SCAN but not in PBE(0)68-D3. }
    \label{sfig:pbe68-r2scan-charges}
\end{figure}



Lastly, we check that the absolute charges are reasonable with respect to DLPNO-CCSD(T) Hirshfeld charges in Fig. \ref{sfig:pbe68-r2scan-charges}. We compare with charges from R2SCAN and observe artificial oxidation for several instances of \ce{Cl-}, indicating persistent self-interaction error. Notably the charges from PBE(0)68-D3 follow well with DLPNO-CCSD(T) and artificial oxidation is not observed.

\subsection{PBE(0)68-D3 vs. DLPNO-CCSD(T) for NEB}
\label{supplemental:pbe-neb-verification}

\begin{figure}
    \centering
    \includegraphics[width=6.5in, center]{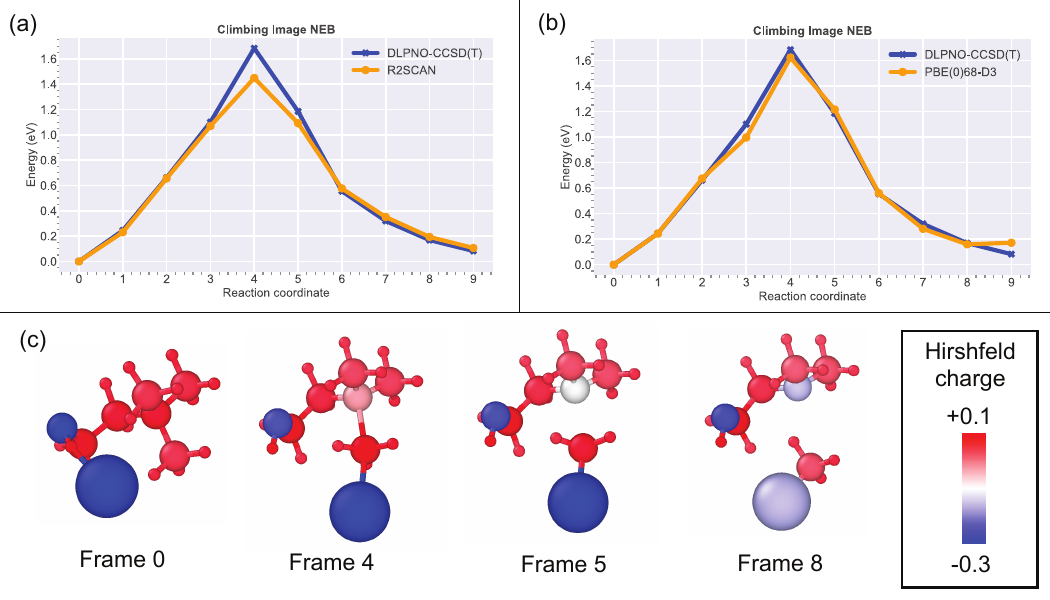}
    \caption{Climbing-Image NEB of ChCl decomposition, calculated in Orca \cite{Neese2020} for two functionals, R2SCAN and PBE(0)68-D3. Hirshfeld charges are colored from low (blue) to high (red).}
    \label{sfig:pbe68-barrier-verify}
\end{figure}

The reliability of PBE(0)68-D3 in predicting the \ce{S_N 2} reaction barrier relative to DLPNO-CCSD(T) is shown in Fig. \ref{sfig:pbe68-barrier-verify}a-b. The Hirshfeld charges colored on each atom during the reaction pathway, are shown for select frames in Fig. \ref{sfig:pbe68-barrier-verify}c. For R2SCAN, the barrier is under-estimated relative to DLPNO-CCSD(T).

Note that the kinetic barrier and thermodynamic difference between products and reactants are larger in bulk solvent: 1) The barrier is +1.68 eV in vacuum (Fig. \ref{sfig:pbe68-barrier-verify}b) compared to bulk solvent of +2 eV. 2) The difference between reactant and product states in vacuum vs. bulk is +7.8 meV/atom (171 meV for 22 atoms of ChCl) and +8.7 meV/atom (1.82 eV for 210 atoms), respectively.

\subsection{Active learning using FLARE workflow}
\label{SI:FLARE}

\begin{figure}
    \centering
    \includegraphics[width=6.5in, center]{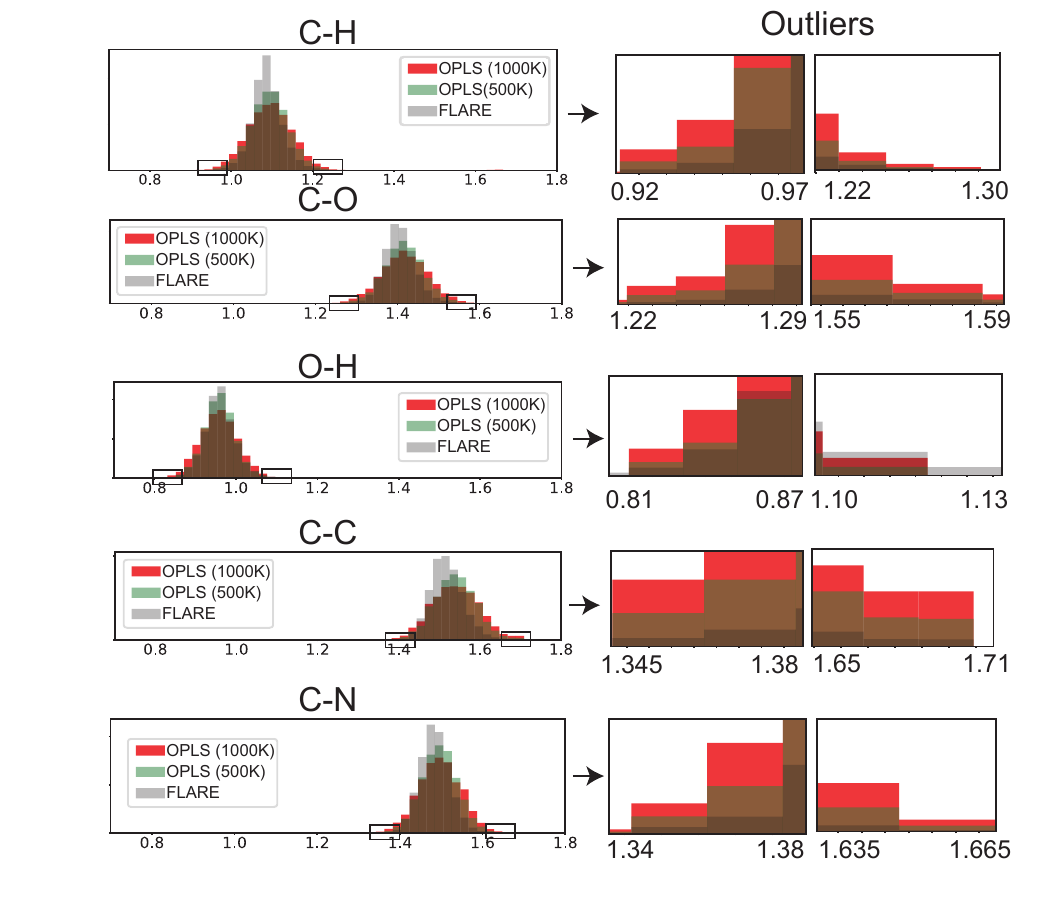}
    \caption{Intramolecular bond lengths sampled from FLARE active learning (grey), compared to those from OPLS at 500 K (green) and 1000 K (red). All colors are partially transparent. Zoomed-in outliers show the overlap in distribution of OPLS (1000K), OPLS (500K), and FLARE.}
    \label{sfig:flare-opls-bl}
\end{figure}

The intramolecular diversity is reflected in Fig. \ref{sfig:flare-opls-bl} by comparing them to OPLS bond lengths at 500 K and 1000 K. 

\begin{figure}
    \centering
    \includegraphics[width=6.5 in, center]{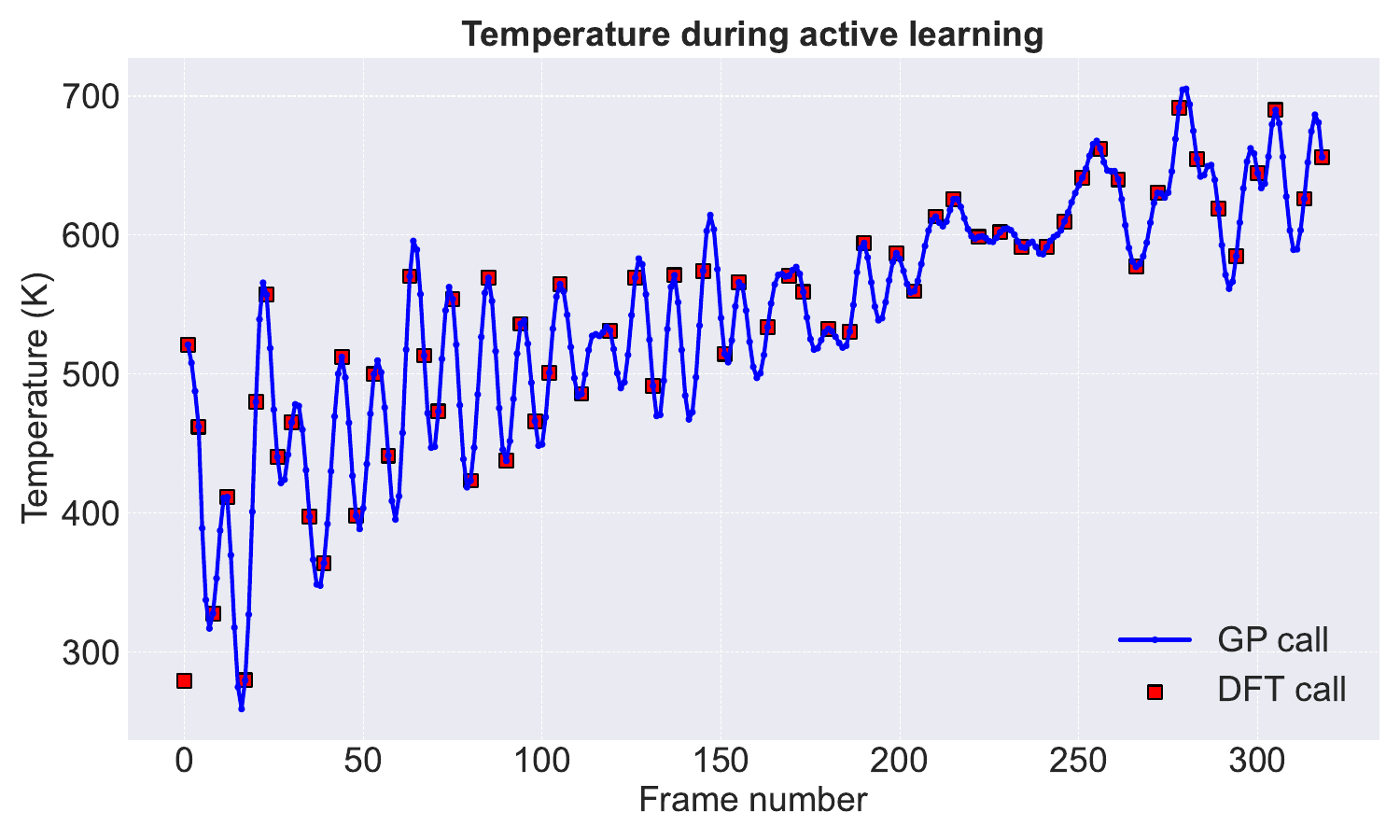}
    \caption{Temperature during active learning trajectory during 0.165 ps of simulation time.}
    \label{sfig:nve-temperature}
\end{figure}

We rationalize the observation in Fig. \ref{sfig:flare-opls-bl} that active learning bond lengths overlap with bond lengths at high temperature classical force fields by examining the temperature of an active learning trajectory in Fig. \ref{sfig:nve-temperature}. Although velocities are initialized to 298 K, the Sparse Gaussian Process continuously explores higher temperatures as new ``uncertain" configurations are added to the sparse set. Thus higher energy bond lengths are automatically explored. 

\subsection{Transition state sampling}

\begin{figure}
    \centering
    \includegraphics[width=6.5 in, center]{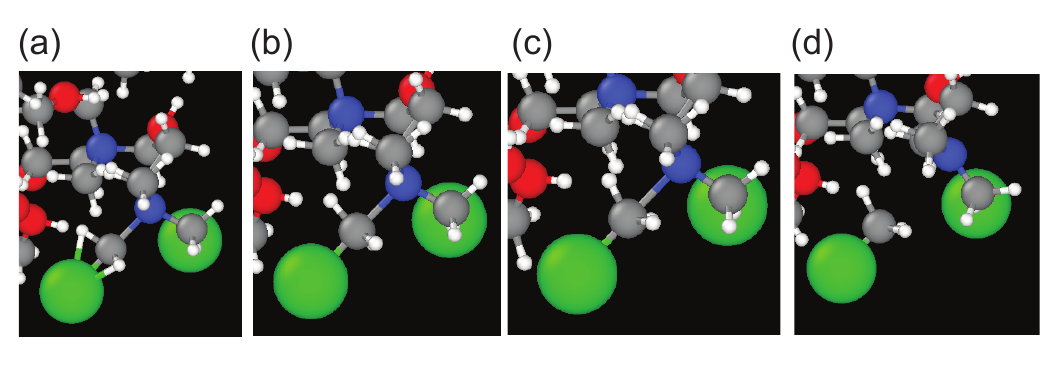}
    \caption{Active learning, beginning from around the transition state. Starting from (a) near the transition state (frame 0/71). (b) Transition state (frame 4/71), followed by (c) inversion (frame 7/71) and (d) chloromethane formation (frame 21/71).}
    \label{sfig:sn2-flare}
\end{figure}

To sample \ce{CH3Cl} formation via the \ce{S_N 2} reaction, we start from a relaxed configuration of 210 atoms, and manually position a nearby \ce{Cl-} exactly along the backside of a \ce{CH3}, 3 \AA\ away. The configuration is not relaxed, but immediately starts a fresh active learning trajectory. All FLARE parameters remain unchanged. There are 71 frames \ce{S_N 2} collected via FLARE active learning, and key snapshots, including the initial frame, are shown in Fig. \ref{sfig:sn2-flare}.

\subsection{Machine learning interatomic potential training}

There are four model test results shown in Tables \ref{stable:allegro-errors} and \ref{stable:allegro-force-errors-by-specie}. The first row corresponds to a model trained only on FLARE active learning (no OPLS frames), and is called ``MLIP-0". The second row, ``MLIP-1", shows results with 100 OPLS frames (300 K and 400 K) added. The third row corresponds to a model additionally trained on the trajectory of MLIP-1 sampled in NVT for 3 ps, and is called ``MLIP-2". The last row is the final model, ``MLIP-3", which is additionally trained on 4 independent Minimum Energy Pathway (MEP) calculations, each calculation with 16 frames. With each model iteration, more data is added to the training.

\begin{table}[h]
\centering
\begin{tabularx}{\textwidth}{c|ccccc}
\toprule
Model & No. test frames & Total frames & \makecell{Energy MAE \\(eV/atom)}& \makecell{Force RMSE\\ (eV/\AA)} & \makecell{Stress RMSE \\(eV/\AA$^3$)}\\
\midrule
MLIP-0& 88 & 592 &$6.8 \times 10^{-5}$& 0.0223& $9.3 \times 10^{-5}$\\ 
MLIP-1& 103 & 692 &$7.8 \times 10^{-5}$& 0.02371& $8.4 \times 10^{-5}$\\ 
MLIP-2 & 112 & 751 & $4.77 \times 10^{-4}$& 0.0707& $3.4 \times 10^{-4}$\\
MLIP-3 & 122 & 815 & $5.39 \times 10^{-4}$& 0.0776& $7.8 \times 10^{-4}$\\
\bottomrule
\end{tabularx}
\caption{Allegro model errors on test set}
\label{stable:allegro-errors}
\end{table}

\begin{table}[h]
\centering
\begin{tabular}{c|ccccc}
\toprule
Model &H & C& O& N & Cl\\
\midrule
MLIP-0& 0.018& 0.023& 0.067& 0.21& 0.019 \\
MLIP-1& 0.0169 & 0.0281 & 0.0397 & 0.0260 & 0.0213\\ 
MLIP-2 & 0.051 &0.105 &0.0677 &0.0836 &0.0598\\ 
MLIP-3 & 0.0604 &0.106 &0.116 &0.087 &0.0679\\
\bottomrule
\end{tabular}
\caption{Allegro model force RMSE on test set, per specie (eV/\AA\ )}
\label{stable:allegro-force-errors-by-specie}
\end{table}

Each iteration of MLIP-$x$ is retrained, drawing from the total number of frames indicated in Table \ref{stable:allegro-errors}. As each re-training is done with a training-validation-testing split of 75\%-15\%-15\%, the number of frames in the test set used to generate the data in Tables \ref{stable:allegro-errors} and \ref{stable:allegro-force-errors-by-specie} are also indicated. 

\subsection{Commentary on MLIP training protocol}
\label{training-discussion}
\begin{figure}
    \centering
    \includegraphics[width=6 in,center]{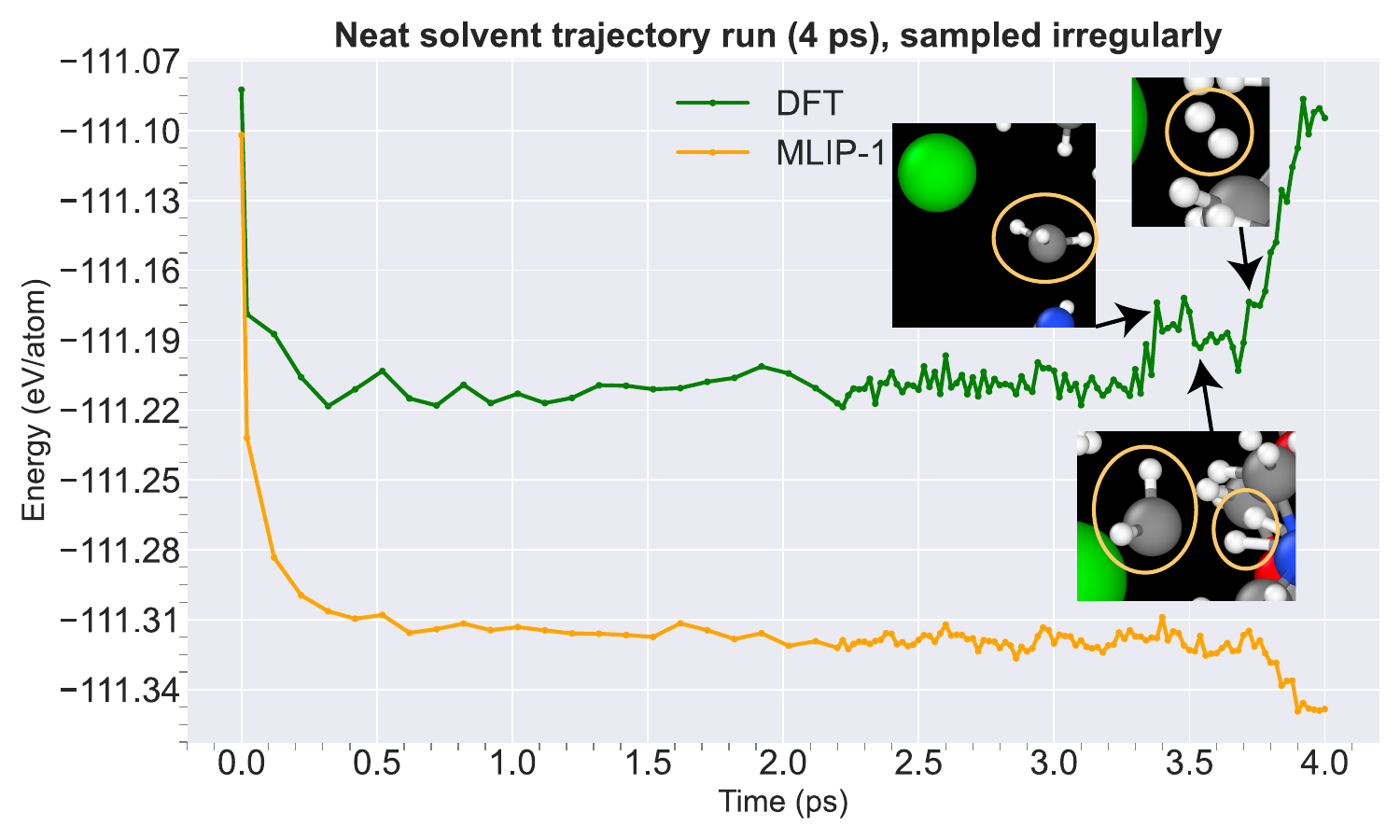}
    \caption{4 ps of MLIP-1 (yellow) with three snapshots of fictitious reactions occur at 3.34, 3.68, and 3.82 ps. Arrows point to the DFT (PBE(0)68-D3) trajectory for ease of understanding that these reactions are high in energy and are not captured by MLIP-1. The formed products in chronological order are: \ce{CH_3} and DMAE; \ce{CH_2} and protonated choline; \ce{H_2} gas, \ce{CH_2)}, and choline. Some of these products are illustrated in the insets (yellow circles). The MLIP-1 predictions show deviation from PBE(0)68-D3 (green), requiring retraining.}
    \label{sfig:allegro-0-4ps}
\end{figure}

\begin{figure}
    \centering
    \includegraphics[width=6 in, center]{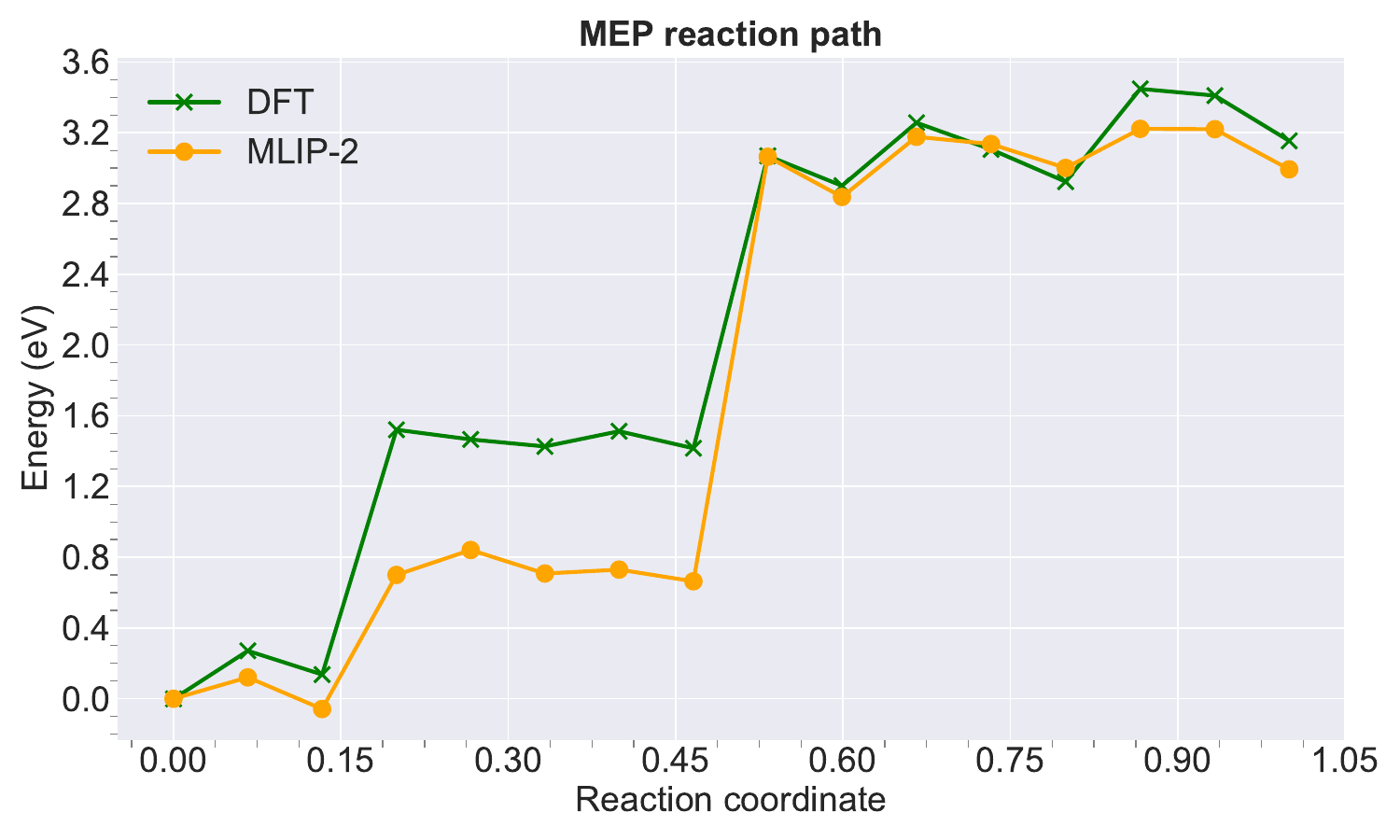}
    \caption{MEP reaction path of MLIP-2 (yellow circles), showing deviation against PBE(0)68-D3 (green x's), requiring retraining.}
    \label{sfig:baseline-mep}
\end{figure}

\begin{figure}
    \centering
    \includegraphics[width=6 in, center]{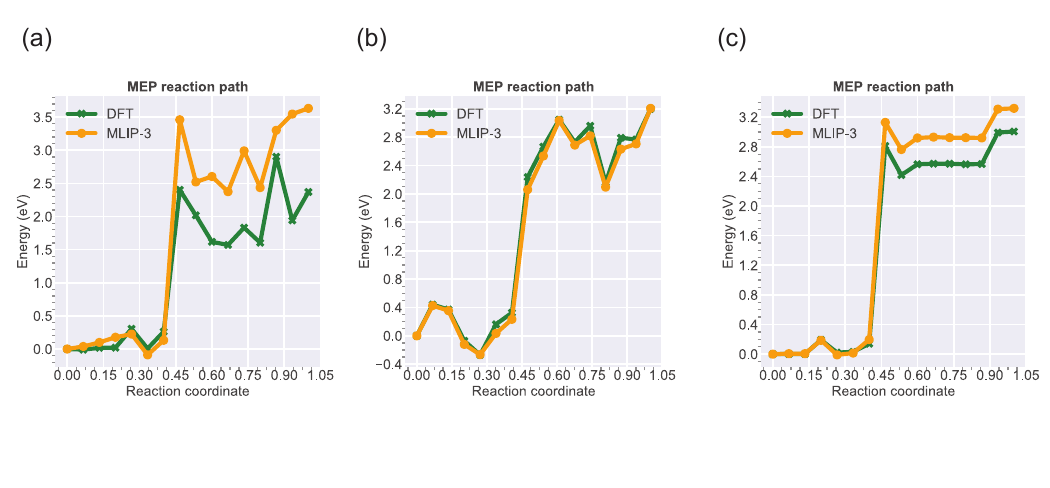}
    \caption{MEP reaction paths of MLIP-3 (yellow circles), showing closer agreement against PBE(0)68-D3 (green x's).}
    \label{sfig:sn2-mep}
\end{figure}

When OPLS frames are not included in the training, deployment of MLIP-0 results in an unstable potential: In NVT at 25 \textdegree C, the system experiences unphysically large forces, throwing a LAMMPS error message in the form of a "lost atom". During the NVT trajectory of MLIP-1 (where OPLS frames included in the training) at 25 \textdegree C, several unphysical reactions occur and there is a large deviation between the MLIP-1 predicted potential energy and the PBE(0)68-D3 energy (Fig. \ref{sfig:allegro-0-4ps}). Further training, using snapshots prior to and during these unphysical reactions, generate a new  ``MLIP-2". NVT sampling of MLIP-2 at 25 \textdegree C shows excellent agreement with PBE(0)68-D3. However, the minimum energy pathway (MEP) calculation (Fig. \ref{sfig:baseline-mep}) shows deviation from PBE(0)68-D3. This mismatch can be corrected by further training, generating ``MLIP-3". The performance of MLIP-3 on three test configurations are in Fig. \ref{sfig:sn2-mep}, indicating reasonable agreement.

\begin{figure}
    \centering
    \includegraphics[width=5 in, center]{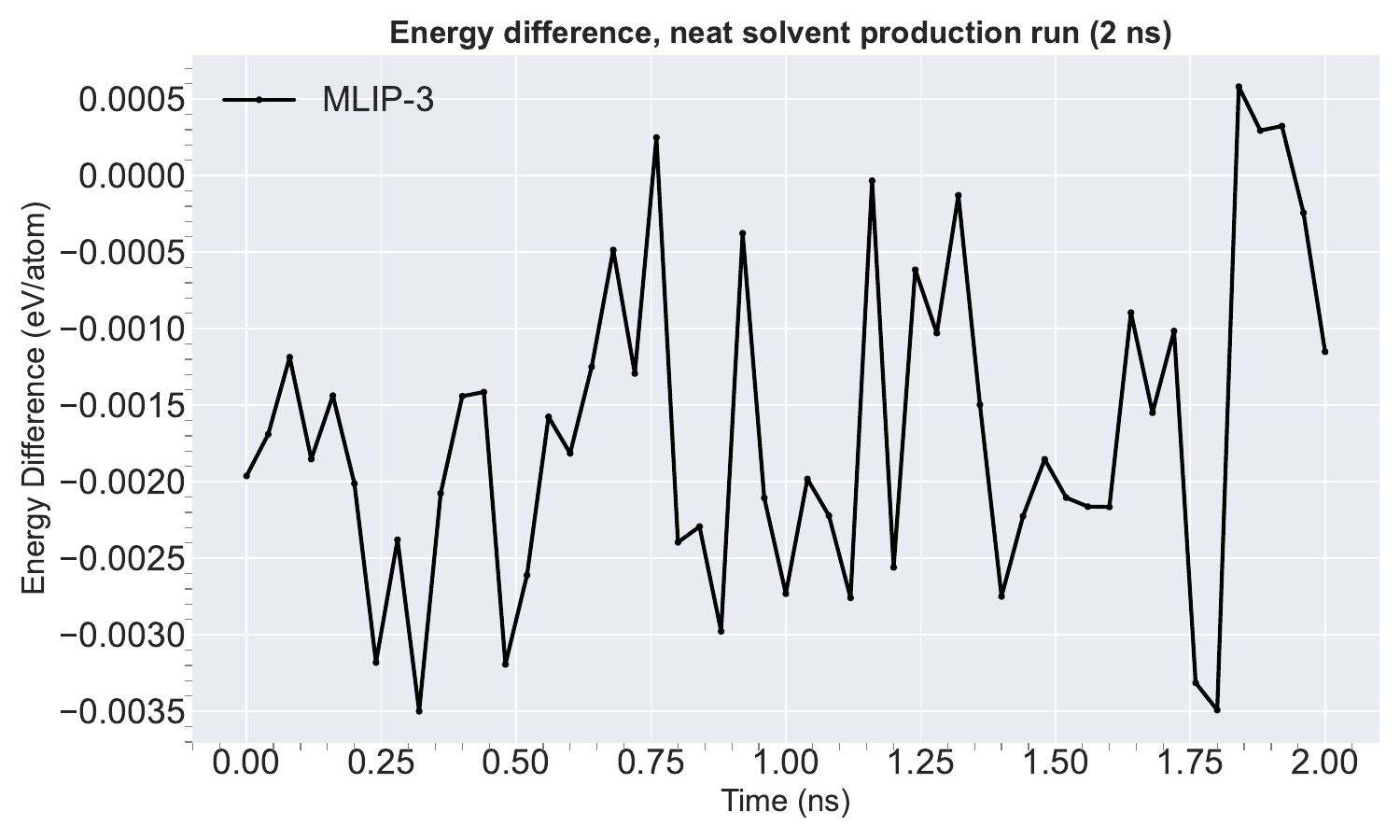}
    \caption{Energy difference (MLIP minus PBE(0)68-D3) during a 2 ns production run of MLIP-3 for neat solvent in NVT at 25 \textdegree C.}
    \label{sfig:nvt-neat-bulk-model}
\end{figure}

\begin{figure}
    \centering
    \includegraphics[width=\textwidth,height=\textheight,keepaspectratio]{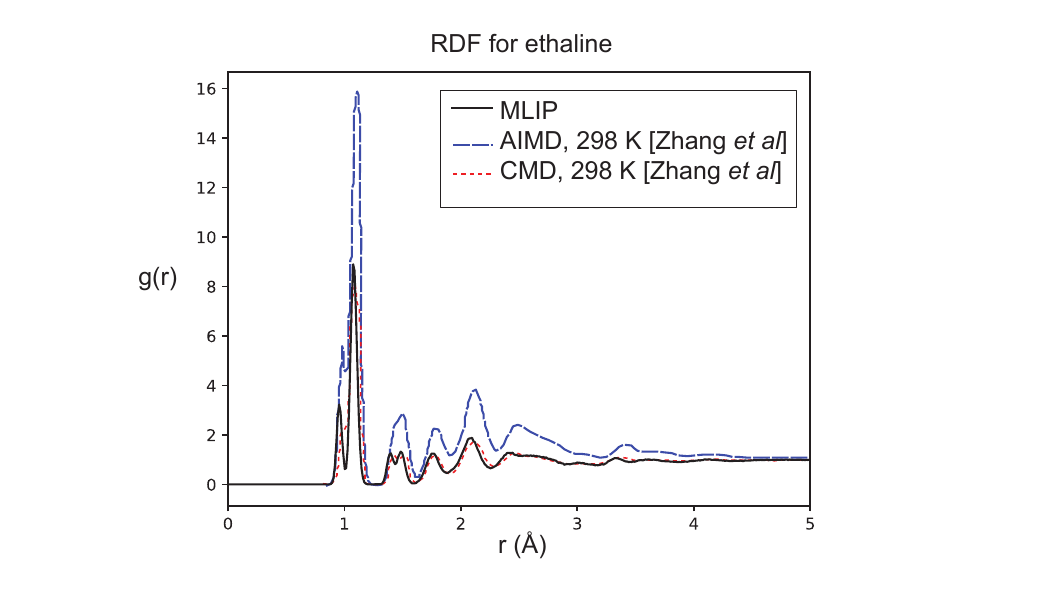}
    \caption{Comparison of RDF of neat ethaline with previously published work by Zhang \textit{et al}\cite{Zhang2020} and MLIP-3.}
    \label{sfig:nvt-rdf-comparison}
\end{figure}

The ``MLIP-3" trajectory in NVT at 25 \textdegree C, checked against PBE(0)68-D3, agree to within 2 meV/atom (Fig. \ref{sfig:nvt-neat-bulk-model}). The bulk liquid structure, radial distribution function (RDF), of ethaline also reproduces previous works (Fig. \ref{sfig:nvt-rdf-comparison}). \cite{Zhang2020}


\clearpage

\subsection{Self-diffusivity calculations}

\begin{figure}
    \centering
    \includegraphics[width=0.4\linewidth]{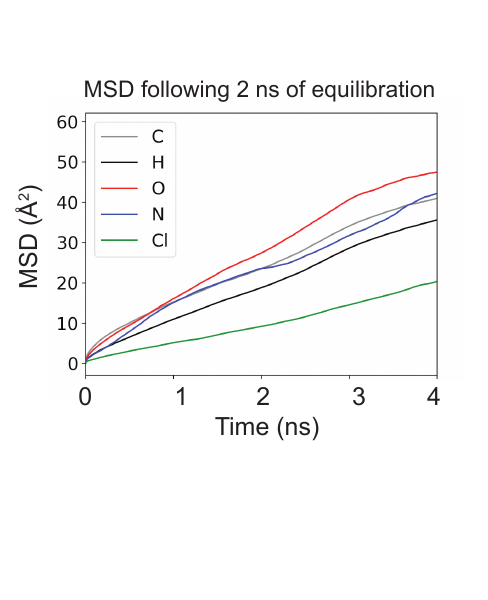}
    \caption{Mean squared displacement (MSD) (\AA$^2$) of C, H, O, N, and Cl calculated by MDAnalysis.\cite{mda1,mda2}}
    \label{sfig:msd}
\end{figure}

The self-diffusivity of EtGl, Ch, and Cl are approximated using the Mean Squared Displacement (MSD) of C, H, O, N, Cl during 4 ns of production after 2 ns of equilibration (Fig. \ref{sfig:msd}). The self-diffusivity for the five species are found by Eqn. \ref{eq:diffn}: 

\begin{equation}
    D=\frac{1}{6}lim_{t\rightarrow\infty}\frac{d}{dt}MSD(r)
    \label{eq:diffn}
\end{equation}

Using the fitted slope from 0-4 ns of Fig. \ref{sfig:msd} for each specie to compute $D$, the diffusivity for C, H, O, N, and Cl, are $1.56\times 10^{-11} m^2/s$, $1.43\times 10^{-11} m^2/s$, $1.96\times 10^{-11} m^2/s$, $1.5\times 10^{-11} m^2/s$, and $0.8\times 10^{-11} m^2/s$, respectively. Since the order of  diffusivity is $D_{Cl}<D_{N}<D_{O}$, we can rank the diffusivity of ChCl and EtGl as $D_{Cl}<D_{Ch}<D_{EtGl}$. Experimentally\cite{mantle} and computationally (based on classical MD force fields fit to choline-chloride-based DESs)\cite{Zhang2020}, $D_{Ch}\approx 1\times 10^{-11}m^2/s$ and $D_{EtGl}\approx 5 \times 10^{-11}m^2/s$.  Thus, the diffusivity of Ch is well-reproduced by the MLIP. 

\clearpage

\subsection{NEB MLIP}

We use MLIP-3 to sample for equilibrium solvation environments during the \ce{S_N 2} reaction, using the nudged elastic band (NEB) method in LAMMPS \cite{Henkelman2000A, Henkelman2000B, Nakano2008,Maras2016}. 

Ethaline is equilibrated for 1 ns in NVT at 25\textdegree C before a production run for 2 ns  under the same conditions. The production energy run tracks well with PBE(0)68-D3 (Fig. \ref{sfig:nvt-neat-bulk-model}). Given the \ce{S_N 2} reaction involves a 180\textdegree alignment of the \ce{Cl-} nucleophile and \ce{CH3} leaving group bonded to N, we also set up the NEB in this manner, starting from environments in the production run where \ce{Cl} and \ce{CH3} are less than 10\textdegree mis-aligned from \ce{C-N} axis and within 6 \AA . In the 2 ns production run, three such environments are obtained and they are sampled at least 400 ps apart. To set up each NEB, we generate (i) an initial frame, aligning \ce{Cl} 180\textdegree\ and 3 \AA\  from \ce{CH3}, and (ii) a final frame, a \ce{CH3Cl} 4.2 \AA\ away from \ce{N}. The energy is minimized, and then 16 frames generate the minimum energy path (MEP). 
The MEP calculation proceeds for 200 ps, following usual protocol. \cite{Henkelman2000A} Note that climbing-image NEB (CI-NEB) was not performed due to unrealistic configurations and numerical instability in LAMMPS, both of which have been observed in other systems\cite{curtin2017, owen2024unbiased}. 

Solvent relaxation is handled by taking each MEP frame after 200 ps, setting all forces on the reacting molecules to 0 (this includes \ce{Cl} and the reacting choline), while equilibrating all other molecules for 200 ps at 25\textdegree\ C. Note that umbrella sampling did not result in physically-reasonable structures near the transition state (TS); neither did setting forces for the reacting \ce{N}, \ce{CH3} and \ce{Cl} to 0. In both cases, this results in the breaking of all three \ce{N-CH3} bonds and formation of isolated \ce{N}. When 126 additional frames from umbrella sampling are added into the training, during umbrella sampling the formation of fictitious \ce{ClClCH_3} is observed instead of \ce{ClCH_3}. While further iterative training of umbrella sampling frames could increase the accuracy of MLIP-x near the TS, we leave this to future work and focus instead on understandings already attainable from the coarse sampling approach.  

\begin{figure}
    \centering
    \includegraphics[width=0.5\linewidth]{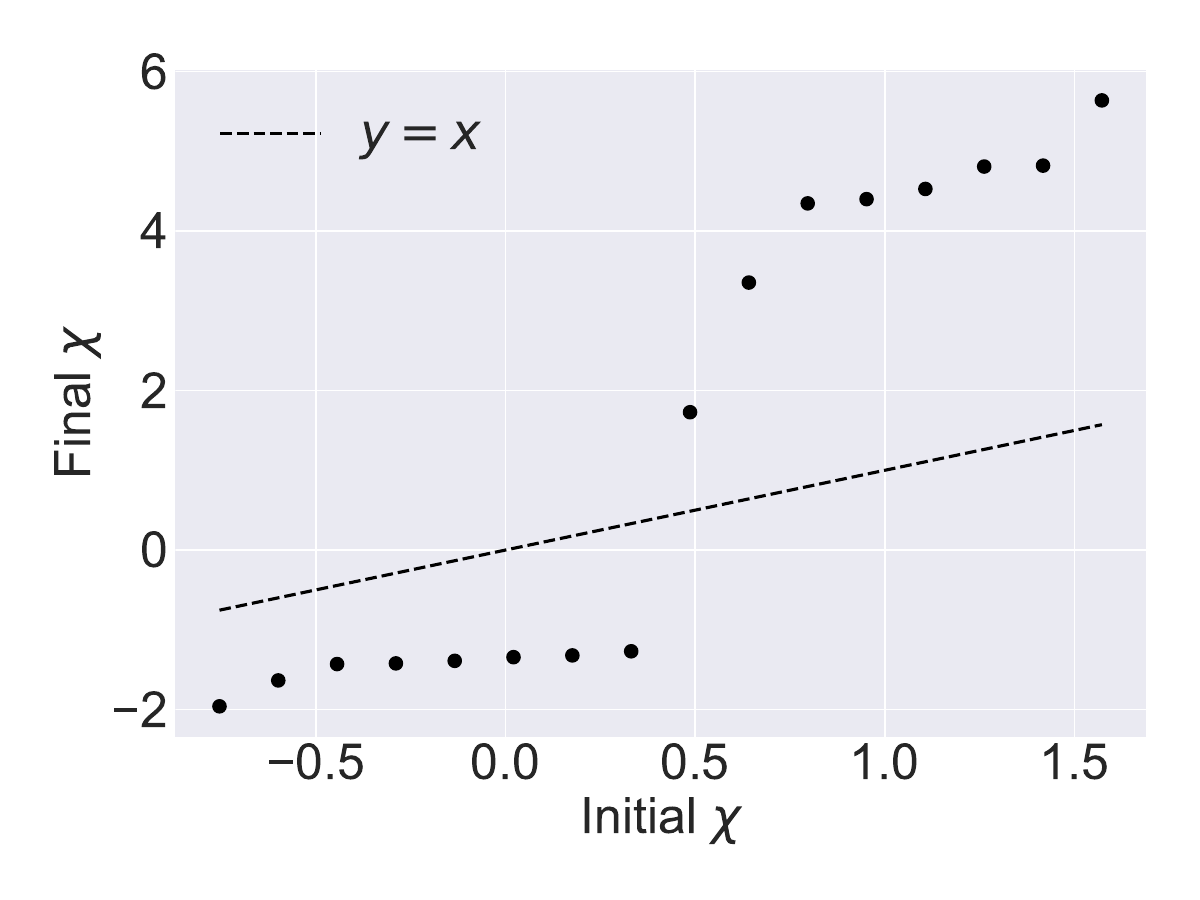}
    \caption{16 MEP images set up the initial set of $\chi$, and they are relaxed by the MLIP to a final set of $\chi$ values. Ideally the final set of $\chi$ are near the initial $\chi$ following $y=x$.}
    \label{sfig:mlip-relaxation-trj}
\end{figure}

The relaxation of MEP calculations away from the TS is summarized in Fig. \ref{sfig:mlip-relaxation-trj}. While there are initial images near the TS ("initial $\chi$"), the final images ("final $\chi$") are moved by the MLIP to $|\chi|\geq 1$. This could be due to the systematic softening observed by Deng \textit{et al} for universal MLIPs \cite{deng2024overcoming}: If the absolute value of the gradient of the potential energy surface is consistently underestimated, this leads to a flattening of the MEP, which drives the system away from the TS. Although non-equilibrium configurations from active learning and iterative training are already included in the training, evidently more high-energy configurations are needed if one desires to sample more finely around the TS. 

For this reason, solvent relaxation is explicitly handled separately from solute relaxation. After 200 ps of solvent equilibration, roughly 500 fs (1000 frames, timestep: 0.5 fs) are simulated in NVE using PBE(0)68-D3. Table \ref{stable:solvent-relax-temp} shows the temperature (std), energy (std) sampled over the NVE MD simulation, along with the MEP reaction coordinate.

\begin{table}[h]
\centering
\begin{tabular}{cccc}
\toprule
$\chi$ & Frames & Temp K (std)& Energy eV (std)\\
\midrule
5.64 &1146& 284.336 (11.0677)&	0	(0.298)\\
4.82 &1140& 291.366	(13.1614)&	0.136	(0.359)\\
4.81 &1141& 287.722	(11.783)&	-0.0272	(0.320)\\
4.53 &1011& 288.977	(11.78)&	0.136	(0.322)\\
4.40 &980& 296.161	(13.654)&	0.136	(0.370)\\
4.35 &1274& 292.131	(11.600)&	0.272	(0.315)\\
3.35&1003& 287.6	(11.851)&	0.571	(0.323)\\
1.72&1007& 278.55	(24.907)&	1.142	(0.771)\\
-1.26&1112& 291.705	(11.637)&	2.067	(0.315)\\
-1.32&1144& 280.716	(11.494)&	2.040	(0.380)\\
-1.34&1145& 281.278	(10.279)&	1.768	(0.282)\\
-1.38&1154& 291.888	(13.836)&	1.986	(0.373)\\
-1.42&1143& 286.484	(11.468)&	1.850	(0.311)\\
-1.43&1135& 295.944	(11.776)&	1.823	(0.320)\\
-1.63&1139& 285.344	(11.477)&	1.523	(0.312)\\
-1.96&1134& 291.568	(11.132)&	1.823	(0.302)\\
\bottomrule
\end{tabular}
\caption{Potential energy and temperature from NVE PBE(0)68-D3 along collective variable $\chi[R]=r_{\ce{Cl-C}}-r_{\ce{N-C}}$. Timestep between frames is 0.5 femtoseconds.}
\label{stable:solvent-relax-temp}
\end{table}

\subsection{CH3Cl + DMAE equilibration}
\begin{figure}
    \centering
    \includegraphics[width=5 in, center]{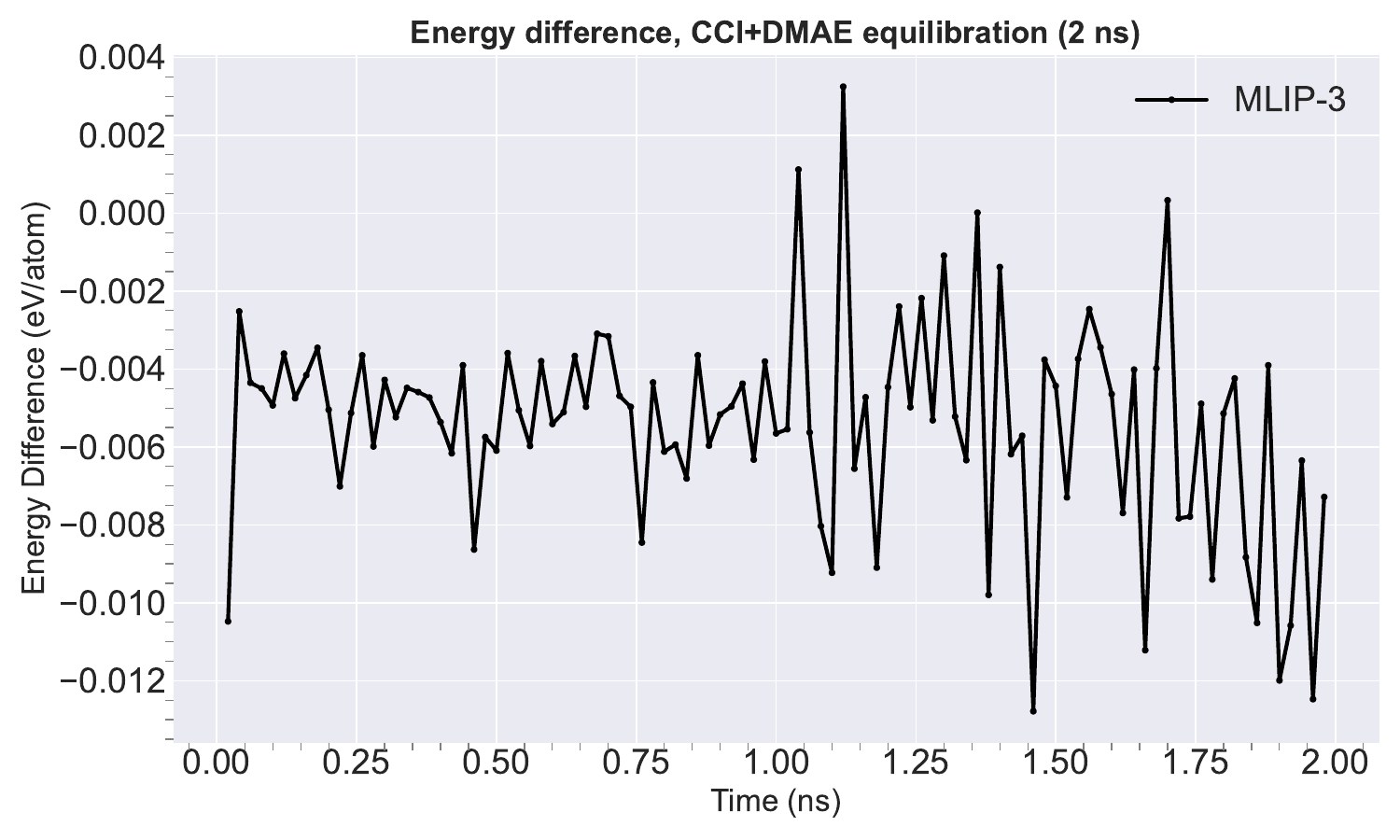}
    \caption{Energy differences (MLIP-3 minus PBE(0)68-D3) for 2 ns.}
    \label{sfig:CH3CL-MLIP-PBE68}
\end{figure}

When \ce{CH3Cl} forms, the solvent energies predicted by the MLIP align well with that of PBE(0)68-D3 for a 2 ns production run (Fig. \ref{sfig:CH3CL-MLIP-PBE68}). 

From the 2 ns production run, the by-specie RDFs for three \ce{Cl} are shown, for \ce{Cl} participating in the reaction, \ce{Cl} near the reaction, and \ce{Cl} further away. The RDF for \ce{N} in DMAE and \ce{N} in choline, are also shown. 

\begin{figure}
    \centering
    \includegraphics[width=6in, center]{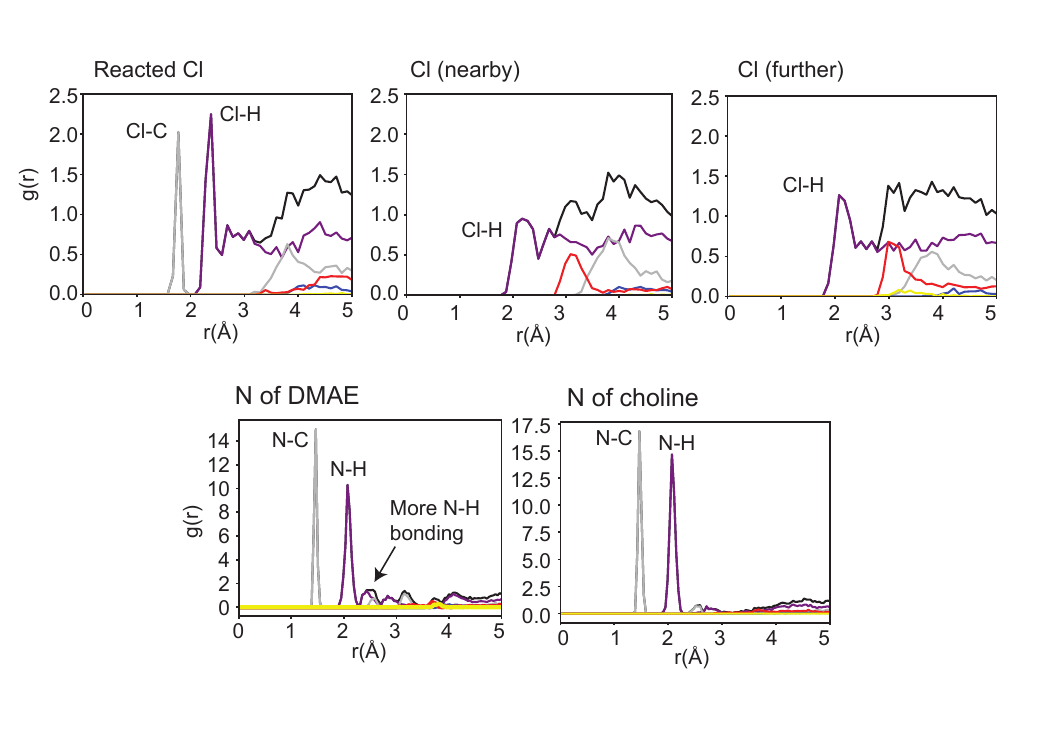}
    \caption{(Top) RDF in partially-reacted solvent for reacted Cl, nearby Cl, and far away Cl from the reaction center. By-specie contributions are colored as: Cl-C (grey), Cl-H (purple), Cl-O (red), Cl-N (blue), Cl-Cl (yellow) and Cl-all (black). (Bottom) RDF in partially-reacted solvent for N in DMAE, and N in choline. By-specie contributions are colored as: N-C (grey), N-H (purple), N-O (red), N-N (blue), N-Cl (yellow), N-all (black).}
    \label{sfig:rdf-all-solutes}
\end{figure}

\clearpage

\section{Experimental}

\subsection{Characterization} 

Water content was measured using a Karl Fisher Titrator (Metrohm 852 Titrando), and averaged over 3 readings for EtGl, ChCl, and Ethaline (Table \ref{stable:water-content}). ChCl was dissolved in MeOH before measurement, and the water content in MeOH was also measured. 

\begin{table}[h]
\centering
\begin{tabular}{ccccc}
\toprule
 & EtGl & ChCl (in EtOH) & EtOH & Ethaline\\
\midrule
1 & 0.08 & 10.04 & 10.51 & 0.57 \\
2 & 0.09 & 10.20 & 10.68 & 0.45 \\ 
3 & 0.06 & 10.58 & 10.51 & 1.25 \\
Average & 0.08 & 10.27 & 10.57 & 0.76 \\
\bottomrule
\end{tabular}
\caption{Water content of EtGl, ChCl, and synthesized ethaline}
\label{stable:water-content}
\end{table}

Thermogravimetric analysis (TGA) of ethaline were conducted using TA Instruments Q500, where the sample was heated at 10 \textdegree C/min from ambient temperature to 1000 \textdegree C under \ch{N2}. 

\begin{figure}
    \centering
\includegraphics[width=6in, center]{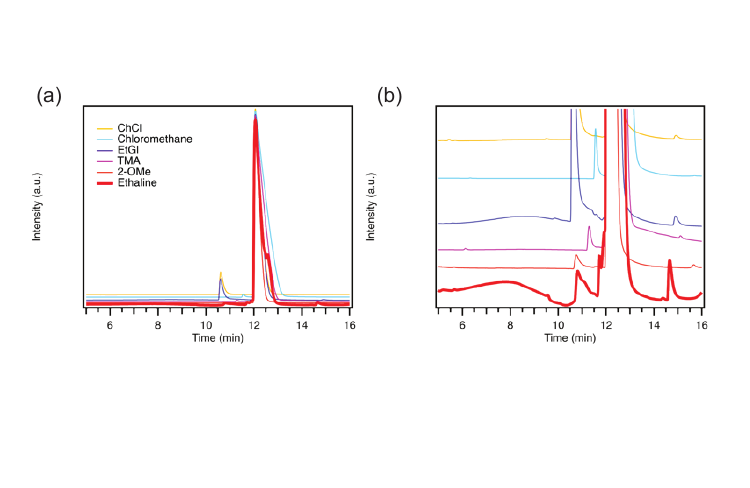}    
\caption{GC-TCD chromatogram of choline chloride (ChCl), chloromethane, ethylene glycol (EtGl), trimethylamine (TMA), 2-methoxymethanol (2-Ome), and ethaline (DES) diluted by MeOH}
    \label{sfig:gc-tcd}
\end{figure}

Analysis of decomposition products after synthesis of ethaline was conducted through gas chromatography (Agilent 7890A), equipped with a HP-PLOT/Q capillary column connected to FID and TCD detectors. GC conditions were as follows: 0.75 min, splitless mode; He, carrier gas; temperature gradient, from 50 to 280 °C at 25°C/min. Peaks from ethaline were benchmarked against standard solutions made from pure DMAE, TMA, dichloromethane, EtGl, ChCl and 2-OMe for identification and quantification (Fig. \ref{sfig:gc-tcd}). Standards and ethaline were prepared by dissolving known concentrations of each chemical in MeOH. 2uL of each solution was drawn using a needle syringe and injected into the GC inlet. Headspace of the synthesized ethaline was also analyzed, and 50uL was taken directly from the vial.

\begin{figure}
    \centering
\includegraphics[width=6in, center]{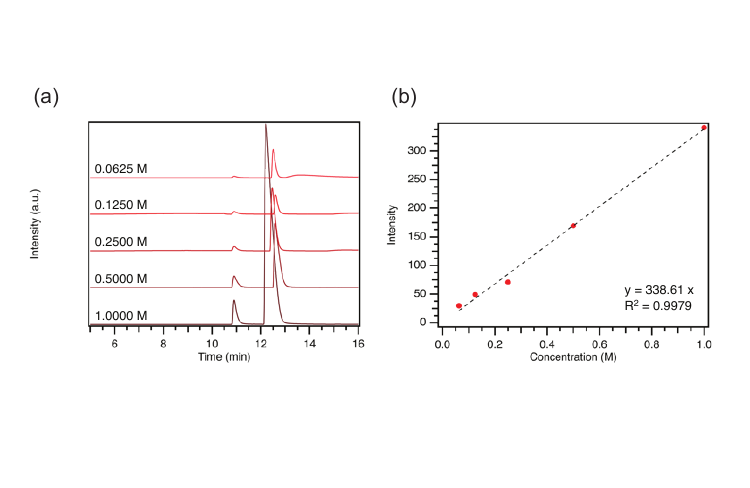}    
\caption{(a) GC-TCD chromatogram of DMAE diluted in MeOH at different concentrations (0.0625 – 1 M) and (b) calibration curve of the intensities at 10.9 min for each concentration.}
    \label{sfig:dmae-conc}
\end{figure}

\clearpage

\section{CP2K sample input file}

\begin{verbatim}
!!! Generated by ASE !!!
&FORCE_EVAL
   METHOD Quickstep
   STRESS_TENSOR ANALYTICAL
   &DFT
      BASIS_SET_FILE_NAME BASIS_ADMM
      BASIS_SET_FILE_NAME BASIS_MOLOPT
      POTENTIAL_FILE_NAME GTH_POTENTIALS
      &MGRID
         REL_CUTOFF 50
         CUTOFF [eV] 6804 
      &END MGRID
      &XC
         &XC_FUNCTIONAL
            &PBE
               SCALE_X 0.3141
               SCALE_C 1.0
            &END PBE
         &END XC_FUNCTIONAL
         &HF
            FRACTION 0.6859
            &SCREENING
               EPS_SCHWARZ 1.0E-6
               SCREEN_ON_INITIAL_P TRUE
            &END SCREENING
            &INTERACTION_POTENTIAL
               POTENTIAL_TYPE TRUNCATED
               CUTOFF_RADIUS 5.0
               T_C_G_DATA t_c_g.dat
            &END INTERACTION_POTENTIAL
            &MEMORY
               MAX_MEMORY 4000
               EPS_STORAGE_SCALING 0.1
            &END MEMORY
         &END HF
         &VDW_POTENTIAL
            DISPERSION_FUNCTIONAL PAIR_POTENTIAL
            &PAIR_POTENTIAL
               TYPE DFTD3
               PARAMETER_FILE_NAME dftd3.dat
               REFERENCE_FUNCTIONAL PBE
               R_CUTOFF [angstrom] 5
            &END PAIR_POTENTIAL
         &END VDW_POTENTIAL
      &END XC
      &SCF
         EPS_SCF 1.0E-6
         MAX_SCF 50
         &OT
            PRECONDITIONER FULL_SINGLE_INVERSE
            MINIMIZER DIIS
         &END OT
         &OUTER_SCF
            MAX_SCF 10
            EPS_SCF 1.0E-6
         &END OUTER_SCF
      &END SCF
      &AUXILIARY_DENSITY_MATRIX_METHOD
         METHOD BASIS_PROJECTION
         ADMM_PURIFICATION_METHOD NONE
      &END AUXILIARY_DENSITY_MATRIX_METHOD
      &LS_SCF
         MAX_SCF 50
      &END LS_SCF
   &END DFT
   &SUBSYS
      &KIND C
         BASIS_SET TZVP-MOLOPT-GTH
         BASIS_SET AUX_FIT pFIT3
         POTENTIAL GTH-PBE
      &END KIND
      &KIND H
         BASIS_SET TZVP-MOLOPT-GTH
         BASIS_SET AUX_FIT pFIT3
         POTENTIAL GTH-PBE
      &END KIND
      &KIND O
         BASIS_SET TZVP-MOLOPT-GTH
         BASIS_SET AUX_FIT pFIT3
         POTENTIAL GTH-PBE
      &END KIND
      &KIND N
         BASIS_SET TZVP-MOLOPT-GTH
         BASIS_SET AUX_FIT pFIT3
         POTENTIAL GTH-PBE
      &END KIND
      &KIND Cl
         BASIS_SET TZVP-MOLOPT-GTH
         BASIS_SET AUX_FIT pFIT3
         POTENTIAL GTH-PBE
      &END KIND
      &COORD
         C 7.045553651703449383e+00 9.093455377147561691e+00 6.125151919623564645e+00
         H 7.149264129369495180e+00 1.005426850705095276e+01 5.621349706944497093e+00
         H 7.468565649406959217e+00 8.386769235430989511e+00 5.393600360731170440e+00
         C 5.623602104657949852e+00 8.862273958478869673e+00 6.496240580623448224e+00
         H 5.090198176081725201e+00 9.714945821458080388e+00 6.828398847498152513e+00
         H 5.750989441840395067e+00 8.143360285968972434e+00 7.360589209283729062e+00
         O 4.853056588578882113e+00 8.146606207009572387e+00 5.517674720166108493e+00
         H 4.358762007442217978e+00 8.843470616862273204e+00 5.083270933211940346e+00
         N 7.966729610322515143e+00 8.908154328182668280e+00 7.283301728122738794e+00
         C 9.354148408913658841e+00 9.423721713514932929e+00 6.944504498812359650e+00
         H 9.866670337331852281e+00 9.400086003841469307e+00 7.920644451088075577e+00
         H 9.821593615312306014e+00 8.731447580733435743e+00 6.210165435513148147e+00
         H 9.407917136828983118e+00 1.047982892124895571e+01 6.680160878493849097e+00
         C 8.157091748544956999e+00 7.488092539580631701e+00 7.564496889777903021e+00
         H 8.505533252330875982e+00 7.041278573136202645e+00 6.574233645628127398e+00
         H 8.983077740133246891e+00 7.418158523585936415e+00 8.334667615842795030e+00
         H 7.317360300626510217e+00 6.934371439827134331e+00 7.766873552825602545e+00
         C 7.458194170368595444e+00 9.753202341865431890e+00 8.394750692399670200e+00
         H 7.037102836613591705e+00 1.067817502603879021e+01 8.065120294016479008e+00
         H 8.212043062633911816e+00 9.875944416028888284e+00 9.196275915622585728e+00
         H 6.704987996410912032e+00 9.186467907652316356e+00 8.888947573939127267e+00
         C 4.557884379085593629e+00 4.767378621544565931e+00 5.835057825515169938e+00
         H 5.495084361282098762e+00 4.519434503733873676e+00 6.095469966475207890e+00
         H 4.514597482135147111e+00 5.868579124303900940e+00 5.352460134452909024e+00
         C 3.911543979052475528e+00 4.708002623558544819e+00 7.146699812133813445e+00
         H 4.106705679544981002e+00 3.687065004686568503e+00 7.589937626417931682e+00
         H 2.870653496736940369e+00 4.935767642387991039e+00 7.129602352123357356e+00
         O 4.519976904138053975e+00 5.679287721677552803e+00 7.981217950634422742e+00
         H 4.253440991919071656e+00 5.547165101366873685e+00 8.929756496184177550e+00
         N 3.901110466150878420e+00 3.869339633040727033e+00 4.880859585073124229e+00
         C 4.075493891096604315e+00 2.423646961488958418e+00 5.257823232539027281e+00
         H 3.305674582834578157e+00 1.892642012375469251e+00 4.585610909150322989e+00
         H 5.001782159760654878e+00 2.070861462804511000e+00 5.110946791929408484e+00
         H 3.784251288642751732e+00 2.217329766302677196e+00 6.266627247272997359e+00
         C 4.573319387843469030e+00 4.051083348667126494e+00 3.546015361621558615e+00
         H 5.604820203552761448e+00 3.672947652852312839e+00 3.589908477091413364e+00
         H 4.027072008498973510e+00 3.578843165796389947e+00 2.850908506217180616e+00
         H 4.543353029889471983e+00 5.147021288398506478e+00 3.344812047182054382e+00
         C 2.524920332651793053e+00 4.146204270085045884e+00 4.780521251244380210e+00
         H 2.029063324866214213e+00 3.800588284936859740e+00 5.683248637765594857e+00
         H 2.079369263149598357e+00 3.649181693297577134e+00 3.919927475911875181e+00
         H 2.416374823651057380e+00 5.232354581952223249e+00 4.944192968867342941e+00
         C 1.232628971109706706e+00 7.472509717772483739e+00 5.604711952751942894e-01
         H 4.316262985618077308e-01 8.098168826621865435e+00 2.208152419837429437e-01
         H 1.877057999103615593e+00 7.276254467531146020e+00 -2.897934396539728619e-01
         C 7.126382097159568163e-01 6.121560663238473587e+00 1.053088608803297577e+00
         H 6.855546544054093028e-01 6.040142930229411178e+00 2.138341702678866696e+00
         H 1.344446181369727933e+00 5.250756139449246795e+00 7.938865892290046800e-01
         O -5.414639865371912997e-01 6.058290046355385705e+00 5.332645126966439886e-01
         H -5.644227399378594079e-01 5.238718448694804408e+00 8.798206319627548122e-02
         N 1.950283874934291006e+00 8.233973224218143372e+00 1.659818118343729054e+00
         C 9.984496608202156898e-01 8.526106654131545781e+00 2.853553719210051742e+00
         H 1.560394248906318504e+00 8.897519219786515166e+00 3.728533637977748505e+00
         H 4.252109423438513169e-01 9.366524946930473661e+00 2.496544379260861479e+00
         H 4.126508878601457542e-01 7.629801897544403744e+00 3.091293623833653470e+00
         C 2.301480941934247237e+00 9.584096561583150375e+00 1.137329760119623945e+00
         H 1.421045919667239588e+00 9.965266691361769347e+00 6.342810790305741797e-01
         H 2.504635008675259122e+00 1.016330172769783324e+01 2.029979784691348765e+00
         H 3.002042089671701053e+00 9.513429489510739856e+00 3.706029622672351587e-01
         C 3.191130017262386875e+00 7.554866964874936919e+00 2.104560548993461566e+00
         H 2.872125788204451524e+00 6.628731371526119354e+00 2.652265060301282507e+00
         H 3.624491951852857330e+00 8.296593066511738002e+00 2.862826768161258251e+00
         H 3.806591858858302757e+00 7.302029419164497881e+00 1.322571190905254213e+00
         C 8.901266654014429847e+00 6.885854178219264377e+00 -7.876703004297190558e-02
         H 9.783724106177738022e+00 6.858436470793590978e+00 -6.678955311976422227e-01
         H 8.335661714945810985e+00 7.722962190286108530e+00 -4.556309831181319381e-01
         C 9.358407570141524801e+00 7.231687203828454535e+00 1.323939813603870697e+00
         H 1.016527921209617702e+01 6.598219850376551676e+00 1.570614070093165449e+00
         H 8.530442589894985161e+00 7.222400614250180872e+00 1.983429579635894946e+00
         O 9.918163111152983547e+00 8.564293593213058386e+00 1.280769962059987055e+00
         H 9.081280528825041998e+00 9.019264171337150415e+00 9.877605144082163280e-01
         N 8.248085040995935202e+00 5.625712520129409810e+00 -3.182110464634647085e-01
         C 9.112678966844473649e+00 4.512984245837755815e+00 4.740335667180562568e-02
         H 8.671128850438355684e+00 3.565289239999805737e+00 -2.351291416851152005e-01
         H 1.000785404675154311e+01 4.646640812897119055e+00 -5.091112857120237978e-01
         H 9.300850276443622278e+00 4.408463822493704498e+00 1.112235772624069252e+00
         C 7.985920451139878473e+00 5.420126199420408852e+00 -1.705865977326234129e+00
         H 8.932814288008163572e+00 5.225286121770621683e+00 -2.218756451603833391e+00
         H 7.346264433409244177e+00 4.544774099050545857e+00 -1.729973038733510604e+00
         H 7.465837194235238172e+00 6.257652312161121166e+00 -2.052693699482050604e+00
         C 7.021740070734828976e+00 5.481452732860090826e+00 4.554595056851262203e-01
         H 7.236171972947186504e+00 5.408304637663240655e+00 1.520133244901717084e+00
         H 6.374618467831268731e+00 4.696652632750001111e+00 5.120133191595502981e-02
         H 6.450276537400697840e+00 6.343411007486804642e+00 3.566546026047999751e-01
         Cl 1.283540279123605110e+01 3.315982789068314673e+00 -6.196120237661352226e-01
         Cl 3.089190096014518438e+00 -1.233871392747174500e+00 4.770635639939417416e+00
         Cl 7.219215589897474139e+00 9.998407969669170114e+00 2.027906548729218605e-01
         Cl 3.883563571669502945e+00 6.120171400343237700e+00 1.092171644391743790e+01
         C 4.441998947410173315e+00 6.793818754035958474e-01 1.656316634362838602e+00
         H 4.585633922178301347e+00 1.497719985510264085e+00 2.432234928401526997e+00
         H 4.437539430495021087e+00 -2.247367581489216060e-01 2.261950648938212183e+00
         C 3.159967420789631554e+00 8.155812915668900764e-01 8.838486026214927849e-01
         H 3.311622995542625070e+00 1.741628259830493963e+00 2.964985193867873958e-01
         H 2.898491732240102259e+00 -8.991270676552162722e-02 1.368048844763289085e-01
         O 2.041403865423534469e+00 9.425128947098156962e-01 1.766577835677187958e+00
         H 1.321988645419661212e+00 1.224822245960056621e+00 1.191631265163059394e+00
         O 5.504996445103055969e+00 7.663152038253937537e-01 8.095659954136712466e-01
         H 5.993404952404691066e+00 -7.519103699011864261e-02 7.162806240835203342e-01
         C 7.186099651947627898e+00 1.113404807793220819e+00 3.414212396467512178e+00
         H 6.940825874841748622e+00 2.074151342821432265e-01 2.883693765859979941e+00
         H 6.442569441126599017e+00 1.215275057520857738e+00 4.281044423095578644e+00
         C 8.609384811117029912e+00 1.014110796658824309e+00 3.970565485294526020e+00
         H 9.354869300898521089e+00 7.059120488422688799e-01 3.180210068298167325e+00
         H 8.744577074105739811e+00 2.056474957767005396e+00 4.337101165018027338e+00
         O 8.892749131796790607e+00 1.272649030531934899e-01 5.056563376937488918e+00
         H 8.450885414820936248e+00 5.520739809860510938e-01 5.867663585471611931e+00
         O 7.059585904565604864e+00 2.197430825565682699e+00 2.556510309928192726e+00
         H 6.715845554528935502e+00 1.863666378129423817e+00 1.730498647157098890e+00
         C 4.384328889780257477e+00 2.785056709740214576e+00 1.007754001946907785e+01
         H 4.657723604387174809e+00 2.884502607160460474e+00 1.109002970310036673e+01
         H 5.158113497558733052e+00 3.205839901279198489e+00 9.515326376601926484e+00
         C 4.244246265613534241e+00 1.297646780538713074e+00 9.668638908352036765e+00
         H 4.964519947120951571e+00 7.011501292021065090e-01 1.029961548143499073e+01
         H 3.249259798308102187e+00 8.307487131559418980e-01 9.960190534811756535e+00
         O 4.576263583132548440e+00 1.234166452319658713e+00 8.240311885632815248e+00
         H 3.880273057339066245e+00 5.177930074005623329e-01 7.934119333488713899e+00
         O 3.179779882016346981e+00 3.407363115633294193e+00 9.697158308385379399e+00
         H 2.972729766461760015e+00 4.160151912194833379e+00 1.022040128050529262e+01
         C 8.754841270161790590e+00 4.360734480249494283e+00 6.116005040733389464e+00
         H 9.706512647983306152e+00 3.854716855377116058e+00 6.063597619828954954e+00
         H 8.677722402865743589e+00 5.170780534525176719e+00 6.891165592167479481e+00
         C 8.561852070931964320e+00 5.085611256330326846e+00 4.835997204897183543e+00
         H 8.922527599438950574e+00 4.419929491441862979e+00 4.106567943363446105e+00
         H 7.509600992375435879e+00 5.341185089361651706e+00 4.776846857662127022e+00
         O 9.385025517810673179e+00 6.266282730425291270e+00 4.811000201946999510e+00
         H 9.776707351889658781e+00 6.366951031511827175e+00 3.975045260201320652e+00
         O 7.718397531600981409e+00 3.490675731369105250e+00 6.343650260135527574e+00
         H 7.823867206523610562e+00 2.918879645478063001e+00 6.967877608145522039e+00
         C 3.207094918129236749e+00 9.171268760935264552e+00 9.557861065330287786e+00
         H 3.121632497091156200e+00 1.017436636670215933e+01 9.241596456987704045e+00
         H 2.379374877404208277e+00 9.052766261498712552e+00 1.018655449042769234e+01
         C 2.900809496325924552e+00 8.314685061161522839e+00 8.299480655024447984e+00
         H 3.638948122101444138e+00 8.288924225560457160e+00 7.483411257152169149e+00
         H 2.767023352869655728e+00 7.319947576139638201e+00 8.475659814406371950e+00
         O 1.596396873761863233e+00 8.711916678124120850e+00 7.793616477096219519e+00
         H 1.660085719698839446e+00 9.476153249526216626e+00 7.244545643528952361e+00
         O 4.451016445493500839e+00 8.972516846501719101e+00 1.010720333814253280e+01
         H 4.563344365011840154e+00 7.998106106394233805e+00 1.029162516254202231e+01
         C -8.310947953129534937e-01 5.352686141847432744e+00 7.699947254143751252e+00
         H -7.871132994335479083e-01 4.340857256801042396e+00 7.410346324869952817e+00
         H -1.791194015157673070e+00 5.767675459291324636e+00 7.498427433460456371e+00
         C -3.837124798936312198e-01 5.416707632727552735e+00 9.139614498718078295e+00
         H 2.828232663603001962e-01 4.652802230619013990e+00 9.389640997242114651e+00
         H 6.301021710204902926e-02 6.418358173722372051e+00 9.362215112103211112e+00
         O -1.547134239561552826e+00 5.140882220427668514e+00 9.825446705061281705e+00
         H -1.298081267530609928e+00 4.589475459245992539e+00 1.055839033087296386e+01
         O 1.798644506878476490e-01 5.989435930314383150e+00 6.967519821893966459e+00
         H 2.721905049714499758e-01 6.895993272852928513e+00 7.079865459545005457e+00
         C 4.955684144388949330e-01 7.850131182794148899e-01 8.865980079918637458e+00
         H 4.887845709724175647e-01 -2.813007187209833337e-01 9.332955630765214394e+00
         H 1.031548065709602335e+00 1.349166562768458366e+00 9.639027863704663801e+00
         C 1.172555082535684523e+00 9.094531117685915600e-01 7.520910792773893405e+00
         H 4.018984015303289636e-01 6.376339174667331466e-01 6.775655510274190441e+00
         H 1.351590328715978284e+00 1.986105012568327499e+00 7.365373401102084827e+00
         O 2.311431601415556258e+00 3.997991085647777804e-02 7.627320980121296223e+00
         H 2.556391433055727802e+00 -2.946119848873503533e-01 6.680485536069888042e+00
         O -7.830267945285405151e-01 1.326472676290698871e+00 8.746796747612622269e+00
         H -8.999006194149951066e-01 2.017047740539368750e+00 9.482160165468910051e+00
         C 7.592002264493484809e+00 1.027314638808376479e+00 9.259141152282042597e+00
         H 7.628024937636666003e+00 1.412320573502440051e-02 9.622844484927972175e+00
         H 6.916855254833335742e+00 1.507498473004555750e+00 9.943510460243144777e+00
         C 9.027854651107819706e+00 1.711045869205141345e+00 9.406418530560848978e+00
         H 9.327879713009778584e+00 1.997516265145278469e+00 1.041051936014165058e+01
         H 9.750510309763210870e+00 9.575077570678477423e-01 9.219327383764950312e+00
         O 9.099895646811329897e+00 2.847492353341430249e+00 8.678461349674849501e+00
         H 9.939482977157480192e+00 2.969266638293285965e+00 8.434131738192023775e+00
         O 7.161202550589372606e+00 1.113919531024439191e+00 7.923751020033908077e+00
         H 6.184799572003876200e+00 1.166603067470253885e+00 7.967293564655610894e+00
      &END COORD
      &CELL
         PERIODIC XYZ
         A 1.300000000000000000e+01 0.000000000000000000e+00 0.000000000000000000e+00
         B 0.000000000000000000e+00 1.181000000000000050e+01 0.000000000000000000e+00
         C 0.000000000000000000e+00 0.000000000000000000e+00 1.199000000000000021e+01
      &END CELL
   &END SUBSYS
   &PRINT
      &STRESS_TENSOR ON
      &END STRESS_TENSOR
      &FORCES ON
      &END FORCES
   &END PRINT
&END FORCE_EVAL
&GLOBAL
   PROJECT cp2k
   PRINT_LEVEL LOW
&END GLOBAL
\end{verbatim}

\clearpage

\bibliography{acs-achemso}